\documentclass[twocolumn]{aastex631}
\usepackage{CJK}
\usepackage{hyperref}

\usepackage{lineno, graphicx}
\newcommand{\Lya}{\hbox{{\rm Ly}$\alpha$}}

\newcommand{\Hb}{\hbox{{\rm H}$\beta$}}

\newcommand{\HeII}{\hbox{{\rm He}\kern 0.1em{\sc ii}}}
\newcommand{\OII}{\hbox{{\rm [O}\kern 0.1em{\sc ii}{\rm ]}}}
\newcommand{\OIII}{\hbox{{\rm [O}\kern 0.1em{\sc iii}{\rm ]}}}
\newcommand{\CIV}{\hbox{{\rm C}\kern 0.1em{\sc iv}}}
\newcommand{\OVI}{\hbox{{\rm O}\kern 0.1em{\sc vi}}}
\newcommand{\NV}{\hbox{{\rm N}\kern 0.1em{\sc v}}}
\newcommand{\PV}{\hbox{{\rm P}\kern 0.1em{\sc v}}}
\newcommand{\CIII}{\hbox{{\rm C}\kern 0.1em{\sc iii}{\rm ]}}}
\newcommand{\MgII}{\hbox{{\rm Mg}\kern 0.1em{\sc ii}}}
\newcommand{\SiIV}{\hbox{{\rm Si}\kern 0.1em{\sc iv}}}
\newcommand{\FeII}{\hbox{{\rm Fe}\kern 0.1em{\sc ii}}}

\newcommand{\pyccf}{\texttt{PyCCF}}

\newcommand{\mrk}{Mrk\,817}
\newcommand{\ngc}{NGC\,5548}

\accepted{December 27, 2023}
\begin{document}
\begin{CJK*}{UTF8}{gbsn}

\title{AGN STORM 2: V. Anomalous Behavior of the \CIV\ Light Curve in \mrk\footnote{Based on observations made with the NASA/ESA Hubble Space Telescope, obtained at the Space Telescope Science Institute, which is operated by the Association of Universities for Research in Astronomy, Inc., under NASA contract NAS5-26555. These observations are associated with program GO-16196.}}

%
%

\author[0000-0002-0957-7151]{Y. Homayouni}
\affiliation{Space Telescope Science Institute, 3700 San Martin Drive, Baltimore, MD 21218, USA}
\affiliation{Department of Astronomy and Astrophysics, The Pennsylvania State University, 525 Davey Laboratory, University Park, PA 16802}
\affiliation{Institute for Gravitation and the Cosmos, The Pennsylvania State University, University Park, PA 16802}

\author[0000-0002-2180-8266]{Gerard A.\ Kriss}
\affiliation{Space Telescope Science Institute, 3700 San Martin Drive, Baltimore, MD 21218, USA}

\author[0000-0003-3242-7052]{Gisella De~Rosa}
\affiliation{Space Telescope Science Institute, 3700 San Martin Drive, Baltimore, MD 21218, USA}

\author[0000-0002-2509-3878]{Rachel Plesha}
\affiliation{Space Telescope Science Institute, 3700 San Martin Drive, Baltimore, MD 21218, USA}

\author[0000-0002-8294-9281]{Edward M.\ Cackett}
\affiliation{Department of Physics and Astronomy, Wayne State University, 666 W.\ Hancock St, Detroit, MI, 48201, USA}

\author[0000-0002-2908-7360]{Michael R.\ Goad}
\affiliation{School of Physics and Astronomy, University of Leicester, University Road, Leicester, LE1 7RH, UK}

\author[0000-0003-0944-1008]{Kirk T.\ Korista}
\affiliation{Department of Physics, Western Michigan University, 1120 Everett Tower, Kalamazoo, MI 49008-5252, USA}

\author[0000-0003-1728-0304]{Keith Horne}
\affiliation{SUPA School of Physics and Astronomy, North Haugh, St.~Andrews, KY16~9SS, Scotland, UK}

\author[0000-0002-3365-8875]{Travis Fischer}
\affiliation{AURA for ESA, Space Telescope Science Institute, 3700 San Martin Drive, Baltimore, MD 21218, USA}

\author[0000-0002-5205-9472]{Tim Waters}
\affiliation{Department of Physics \& Astronomy, 
University of Nevada, Las Vegas 
4505 S. Maryland Pkwy, 
Las Vegas, NV, 89154-4002, USA}

\author[0000-0002-3026-0562]{Aaron J.\ Barth}
\affiliation{Department of Physics and Astronomy, 4129 Frederick Reines Hall, University of California, Irvine, CA, 92697-4575, USA}

\author[0000-0003-0172-0854]{Erin A.\ Kara}
\affiliation{MIT Kavli Institute for Astrophysics and Space Research, Massachusetts Institute of Technology, Cambridge, MA 02139, USA}

\author[0000-0001-8391-6900]{Hermine Landt}
\affiliation{Centre for Extragalactic Astronomy, Department of Physics, Durham University, South Road, Durham DH1 3LE, UK}

\author[0000-0003-2991-4618]{Nahum Arav}
\affiliation{Department of Physics, Virginia Tech, Blacksburg, VA 24061, USA}

\author[0000-0001-6301-570X]{Benjamin D. Boizelle}
\affiliation{Department of Physics and Astronomy, N284 ESC, Brigham Young University, Provo, UT, 84602, USA}

\author[0000-0002-2816-5398]{Misty C.\ Bentz}
\affiliation{Department of Physics and Astronomy, Georgia State University, 25 Park Place, Suite 605, Atlanta, GA 30303, USA}


\author[0000-0002-1207-0909]{Michael S.\ Brotherton}
\affiliation{Department of Physics and Astronomy, University of Wyoming, Laramie, WY 82071, USA}

\author[0000-0002-4830-7787]{Doron Chelouche}
\affiliation{Department of Physics, Faculty of Natural Sciences, University of Haifa, Haifa 3498838, Israel}
\affiliation{Haifa Research Center for Theoretical Physics and Astrophysics, University of Haifa, Haifa 3498838, Israel}

\author[0000-0001-9931-8681]{Elena Dalla Bont\`{a}}
\affiliation{Dipartimento di Fisica e Astronomia ``G.\  Galilei,'' Universit\`{a} di Padova, Vicolo dell'Osservatorio 3, I-35122 Padova, Italy}
\affiliation{INAF-Osservatorio Astronomico di Padova, Vicolo dell'Osservatorio 5 I-35122, Padova, Italy}

\author[0000-0002-0964-7500]{Maryam Dehghanian}
\affiliation{Department of Physics and Astronomy, The University of Kentucky, Lexington, KY 40506, USA}

\author[0000-0002-5830-3544]{Pu Du} 
\affiliation{Key Laboratory for Particle Astrophysics, Institute of High Energy Physics, Chinese Academy of Sciences, 19B Yuquan Road,\\ Beijing 100049, People's Republic of China}

\author[0000-0003-4503-6333]{Gary J.\ Ferland}
\affiliation{Department of Physics and Astronomy, The University of Kentucky, Lexington, KY 40506, USA}


\author[0000-0002-2306-9372]{Carina Fian}
\affiliation{Haifa Research Center for Theoretical Physics and Astrophysics, University of Haifa, Haifa 3498838, Israel}
\affiliation{School of Physics and Astronomy and Wise observatory, Tel Aviv University, Tel Aviv 6997801, Israel}



\author[0000-0001-9092-8619]{Jonathan Gelbord}
\affiliation{Spectral Sciences Inc., 4 Fourth Ave., Burlington, MA 01803, USA}



\author[0000-0001-9920-6057]{Catherine J. Grier}
\affiliation{Department of Astronomy, University of Wisconsin-Madison, Madison, WI 53706, USA} 

\author[0000-0002-1763-5825]{Patrick B.\ Hall}
\affiliation{Department of Physics and Astronomy, York University, Toronto, ON M3J 1P3, Canada}


\author{Chen Hu}
\affiliation{Key Laboratory for Particle Astrophysics, Institute of High Energy Physics, Chinese Academy of Sciences, 19B Yuquan Road, Beijing 100049, People's Republic of China}

\author[0000-0002-1134-4015]{Dragana Ili\'{c}}
\affiliation{University of Belgrade - Faculty of Mathematics, Department of astronomy, Studentski trg 16 Belgrade, Serbia}
\affiliation{Hamburger Sternwarte, Universit{\"a}t Hamburg, Gojenbergsweg 112, 21029 Hamburg, Germany}

\author[0000-0003-0634-8449]{Michael D.\ Joner}
\affiliation{Department of Physics and Astronomy, N284 ESC, Brigham Young University, Provo, UT, 84602, USA}

\author[0000-0001-5540-2822]{Jelle Kaastra}
\affiliation{SRON Netherlands Institute for Space Research, Niels Bohrweg 4, 2333 CA Leiden, The Netherlands}
\affiliation{Leiden Observatory, Leiden University, PO Box 9513, 2300 RA Leiden, The Netherlands}

\author[0000-0002-9925-534X]{Shai Kaspi}
\affiliation{School of Physics and Astronomy and Wise observatory, Tel Aviv University, Tel Aviv 6997801, Israel}

\author[0000-0001-5139-1978]{Andjelka B. Kova{\v c}evi{\'c}}
\affiliation{University of Belgrade - Faculty of Mathematics, Department of astronomy, Studentski trg 16 Belgrade, Serbia}

\author[0000-0001-8638-3687]{Daniel Kynoch}
\affiliation{School of Physics and Astronomy, University of Southampton, Highfield, Southampton SO17 1BJ, UK}
\affiliation{Astronomical Institute of the Czech Academy of Sciences, Bo\v{c}n\'{i} II 1401, 141 00 Prague, Czechia}

\author[0000-0001-5841-9179]{Yan-Rong Li}
\affiliation{Key Laboratory for Particle Astrophysics, Institute of High Energy Physics, Chinese Academy of Sciences, 19B Yuquan Road,\\ Beijing 100049, People's Republic of China}



\author[0000-0002-4992-4664]{Missagh Mehdipour}
\affiliation{Space Telescope Science Institute, 3700 San Martin Drive, Baltimore, MD 21218, USA}

\author[0000-0001-8475-8027]{Jake A. Miller}
\affiliation{Department of Physics and Astronomy, Wayne State University, 666 W.\ Hancock St, Detroit, MI, 48201, USA}

\author{Jake Mitchell}
\affiliation{Centre for Extragalactic Astronomy, Department of Physics, Durham University, South Road, Durham DH1 3LE, UK}

\author[0000-0001-5639-5484]{John Montano}
\affiliation{Department of Physics and Astronomy, 4129 Frederick Reines Hall, University of California, Irvine, CA, 92697-4575, USA}

\author[0000-0002-6766-0260]{Hagai Netzer}
\affiliation{School of Physics and Astronomy and Wise observatory, Tel Aviv University, Tel Aviv 6997801, Israel}

\author[0000-0001-7351-2531]{J. M. M. Neustadt}
\affiliation{Department of Astronomy, The Ohio State University, 140 West 18th Avenue, Columbus OH 43210}



\author[0000-0003-1183-1574]{Ethan Partington}
\affiliation{Department of Physics and Astronomy, Wayne State University, 666 W.\ Hancock St, Detroit, MI, 48201, USA}


\author[0000-0003-2398-7664]{Luka \v{C}.\ Popovi\'{c}}
\affiliation{Astronomical Observatory, Volgina 7, 11060 Belgrade, Serbia}
\affiliation{University of Belgrade - Faculty of Mathematics, Department of astronomy, Studentski trg 16 Belgrade, Serbia}

\author[0000-0002-6336-5125]{Daniel Proga}
\affiliation{Department of Physics \& Astronomy, 
University of Nevada, Las Vegas 
4505 S.\ Maryland Pkwy, 
Las Vegas, NV, 89154-4002, USA}



\author[0000-0003-1772-0023]{Thaisa Storchi-Bergmann}
\affiliation{Departamento de Astronomia - IF, Universidade Federal do Rio Grande do Sul, CP 150501, 91501-970 Porto Alegre, RS, Brazil}

\author[0000-0002-9238-9521]{David Sanmartim}
\affiliation{Rubin Observatory Project Office, 950 N. Cherry Ave., Tucson, AZ 85719, USA} 

\author[0000-0003-2445-3891]{Matthew R.\ Siebert}
\affiliation{Department of Astronomy and Astrophysics, University of California, Santa Cruz, CA 92064, USA}

\author[0000-0002-8460-0390]{Tommaso Treu}\thanks{Packard Fellow}
\affiliation{Department of Physics and Astronomy, University of California, Los Angeles, CA 90095, USA}


\author[0000-0001-9191-9837]{Marianne Vestergaard}
\affiliation{Steward Observatory, University of Arizona, 933 North Cherry Avenue, Tucson, AZ 85721, USA}
\affiliation{DARK, The Niels Bohr Institute, University of Copenhagen, Jagtvej 155, DK-2200 Copenhagen, Denmark}

\author[0000-0001-9449-9268]{Jian-Min Wang}
\affiliation{Key Laboratory for Particle Astrophysics, Institute of High Energy Physics, Chinese Academy of Sciences, 19B Yuquan Road,\\ Beijing 100049, People's Republic of China}
\affiliation{School of Astronomy and Space Sciences, University of Chinese Academy of Sciences, 19A Yuquan Road, Beijing 100049, People's Republic of China}
\affiliation{National Astronomical Observatories of China, 20A Datun Road, Beijing 100020, People's Republic of China}

\author[0000-0003-1810-0889]{Martin J.\ Ward}
\affiliation{Centre for Extragalactic Astronomy, Department of Physics, Durham University, South Road, Durham DH1 3LE, UK}



\author[0000-0003-0931-0868 ]{Fatima Zaidouni}
\affiliation{MIT Kavli Institute for Astrophysics and Space Research, Massachusetts Institute of Technology, Cambridge, MA 02139, USA}

\author[0000-0001-6966-6925]{Ying Zu}
\affiliation{Department of Astronomy, School of Physics and Astronomy, Shanghai Jiao Tong University, 800 Dongchuan Road, Shanghai, 200240, People's Republic of China}
\affiliation{Shanghai Key Laboratory for Particle Physics and Cosmology, Shanghai Jiao Tong University, Shanghai 200240, People's Republic of China}









\begin{abstract}
An intensive reverberation mapping campaign on the Seyfert 1 galaxy \mrk\
using the Cosmic Origins Spectrograph (COS) on the Hubble Space Telescope (HST) revealed significant variations in the response of the broad UV emission lines to fluctuations in the continuum emission.
The response of the prominent UV emission lines changes over a $\sim$60-day duration, resulting in
distinctly different time lags in the various segments of the light curve over the 14 months observing campaign. One-dimensional echo-mapping models fit these variations if a slowly varying background is included for each emission line. These variations are more evident in the \ion{C}{4} light curve, which is the line least affected by intrinsic absorption in \mrk\ and least blended with neighboring emission
lines. We identify five temporal windows with distinct emission line response, and measure their corresponding time delays, which range from 2 to 13 days.
These temporal windows are plausibly linked to changes in the UV and X-ray obscuration occurring during these same intervals.
The shortest time lags occur during periods with diminishing obscuration,
whereas the longest lags occur during periods with rising obscuration.
We propose that the obscuring outflow shields the ultraviolet broad lines from the ionizing continuum. The resulting change in the spectral energy distribution of the ionizing continuum, as seen by clouds at a range of distances from the nucleus, is responsible for the changes in the line response.
\end{abstract}

\section{Introduction}\label{sec:intro}
The broad emission-line regions (BLR) are of paramount importance to the study of active galaxy nuclei (AGN) as they provide a probe of the central regions of AGN and their physical conditions.
It has long been established that photoionization by the nuclear continuum is responsible for driving the observed ultraviolet (UV) emission lines \citep{Krolik1999}. Models predict that photoionization heats the broad-line gas, and that much of the \ion{C}{4} emission is due to collisional excitation processes. The observed far-UV continuum, i.e., the closest wavelength window to the ionizing continuum, is only a proxy for the ionizing continuum ($\lambda \leqslant\,$912\,\AA) that is generally unobservable due to the Lyman limit of our own Galaxy and the presence of hydrogen in the AGN host galaxy. Changes in the ionizing continuum flux from the central source lead to correlated changes in the broad emission lines produced in the BLR. Non-linear responses in the broad emission-line fluxes can be caused by a mixture of BLR clouds with a range of column densities and ionization parameters. Additionally, temporal changes in the ionizing flux with a fixed spectral shape will result in non-linear changes in the emission line flux for most emission lines \citep{Goad2004, Goad2015}.
Intrinsic to each object, this nonlinear correlation is nominally referred to as the ``intrinsic Baldwin effect" \citep{Kinney1990, krolik1991, Pogge1992, Goad2004}. The complex relationship between the continuum flux, $F_{\rm continuum}$, and the emission-line flux, $F_{\rm line}$, can be parameterize by two factors: the reprocessing efficiency and the marginal response. The reprocessing efficiency for some particular emission line is the fraction of incident ionizing photons reprocessed into that emission line, and is related to the equivalent width (EW) of the emission line. Here, this EW is determined relative to our proxy for the time-variable strength of the incident ionizing continuum flux, the flux at 1180~\AA.
The marginal response of an emission line is a measure of how the reprocessing efficiency changes as a function of the strength of the driving ionizing continuum. This relation is parameterized by $F_{\rm line}\propto F_{\rm continuum}^{\eta_{\,\rm eff}}$, where $\eta_{\rm eff}$ is a measure of the instantaneous emission-line response to the ionizing continuum variations and is typically measured after first removing non-variable background contamination (e.g., narrow emission lines and host galaxy contribution), and after correcting for the mean delay between the continuum and emission line variations. Here, again, we use the strength of the continuum at 1180~\AA\ as a proxy for the strength of the largely unobserved ionizing continuum.
The marginal response of an emission line is generally calculated as the logarithmic slope of the $F_{\rm line}$ vs. $F_{\rm continuum}$ relation. 
In terms of equivalent width (EW), this relationship can be expressed as $\rm EW_{\rm line}\propto$ $F_{\rm continuum}^{\beta}$, with $\beta = \eta_{\rm eff} - 1$.
Generally, the emission-line response to the continuum variations is weaker than linear, so that $\eta_{\rm eff}<1$ (i.e., $\beta<0$) \citep{Pogge1992, Gilbert2003, Goad2004, Goad2014, Goad2016}. 
This relationship can also explain some of the scatter in the global Baldwin effect, where observations among different AGN exhibit an anti-correlation between emission-line EW and the continuum level \citep{Baldwin1977, Kinney1990, Osmer1994, Cackett2006}. In this work, we refer to the slope in $F_{\rm line}$ vs. $F_{\rm continuum}$ relation as the marginal response, and the normalization is connected to the line EW, which characterizes the reprocessing efficiency.

Over the past three decades, reverberation mapping (RM; \citealp{Blandford1982, Peterson1993, Peterson2004}) has been a successful technique for mapping the inner structure of AGN. The RM technique relies on the following assumptions: (a) the central ionizing source is point-like (b) the AGN variability at two different wavelengths is causally connected (c) the light travel time is the most important timescale. With AGN continua showing variability on timescales of days to years, the time delay between the fluctuations in the continuum and the emission-line response is believed to be a measure of the mean physical distance between the continuum emitting region around supermassive black hole (SMBH) and the BLR, assuming that the photons travel freely. Assuming that the BLR-gas motion is primarily gravitational and dominates the velocity dispersion of BLR gas, the virial product then provides a means of measuring the SMBH mass and studying the  structure of the BLR (\citealp{Clavel1991, Peterson1991, Horne1991, Kaspi2000, Peterson2004, Bentz2009, Grier2013, Pancoast2014, Barth2015, Du2016, Pei2017, Du2018, DeRosa2018, Brotherton2020, Bentz2021, U2022, Bao2022}). An extensive review of RM applied at a range of wavelengths has recently been published by \citet{Cackett2021}. 

To probe the spatial and kinematic structure of the BLR gas, a one-dimensional (1D) description of the BLR response function is not sufficient \citep{Welsh1991}. A more complete form of RM is ``velocity-resolved" RM \citep{Bahcall1972, Blandford1982}, which measures the projection of the BLR into two observables, the line-of-sight velocity and the time delay response of the BLR. The 2D velocity-delay map encodes information about the BLR geometry and kinematics \citep{krolik1991, Ulrich1996, Bentz2010a, Barth2011, Pancoast2011,Grier2013,Pancoast2014,Du2016,Pei2017,Li2018,Bentz2021,Horne2021,U2022,Villafana2022}.
Despite decades of RM observation and several optical velocity-resolved RM campaigns, only one velocity-resolved RM campaign in the UV with HST has been conducted \citep{DeRosa2015}, and the only other UV velocity-resolved campaign on NGC~4151 was based on IUE monitoring \citep{Ulrich1996}. This is due to the demanding nature of velocity-resolved RM in terms of data quality, time resolution, and duration \citep{Horne2004}. The Space Telescope and Optical Reverberation Mapping (STORM; \citealp{DeRosa2015, Kriss2019b}) project used daily observations of NGC 5548 over six months with the Cosmic Origins Spectrograph (COS; \citealp{Green2012}) on the Hubble Space Telescope (HST) to carry out velocity-resolved RM. The AGN STORM program (hereafter referred to as AGN STORM~1) was accompanied by near-UV and X-ray monitoring with Swift \citep{Edelson2015, Fausnaugh2016}, optical ground-based spectroscopy \citep{Pei2017}, and four X-ray observations with Chandra \citep{Mathur2017}. AGN STORM~2 is a second such program targetting the Seyfert~1 galaxy \mrk\ (z = 0.03146, $\lambda L_{\lambda 5100}$ = 43.78 $\rm erg\,s^{-1}$, $M_{\rm BH} = 3.86^{+0.61}_{-0.59}\times 10^{7}M_{\odot}$ \citep{Peterson1998, Peterson2004, Denney2010, Bentz2013} with intensive multi-wavelength monitoring (see Section \ref{sec:data} for more detail).

The AGN STORM~1 observations have unveiled a wealth of information about the structure of the BLR, the accretion disk, and the associated outflowing winds.
One of the most unexpected results of the AGN STORM~1 program was that approximately midway into the campaign, the emission lines decorrelated from the continuum fluctuations as manifested in a sudden and sustained drop in the emission-line flux and EW and an apparent lack of response to the continuum flux variations during that time period. The lines remained decorrelated for 65-70 days, but became well-correlated again at the end of the campaign \citep{Goad2016}. During this anomalous period, the emission-line response amplitude is also significantly lower compared to the observed continuum variations. Similar changes occurred in the high-ionization narrow absorption lines. Both effects may be induced by the presence of outflows that obscure the ionizing flux, resulting in ``BLR holidays" \citep{Dehghanian2019a}, where the emission lines become weaker and their variations less correlated with those in the continuum \citep{Goad2016}. This implies that the simple reverberation-mapping picture, which relies on continuum/emission-line reverberation, is far more complex than we originally anticipated. 

The primary goal of the AGN STORM~2 program is to study a second AGN with intensive multi-wavelength monitoring. Although \mrk\ was selected from spectra taken in 2009 to avoid the X-ray and UV obscuration that complicated the STORM~1 campaign, the first COS spectra of \mrk\ showed the presence of strong, broad, blue-shifted UV absorption troughs similar to the obscuring outflows seen in many other Seyfert galaxies (Mrk\,335; \citealp{Longinotti2013, Longinotti2019, Parker2019}, \ngc; \citealp{Kaastra2014}, NGC\,985; \citealp{Ebrero2016}, NGC\,3783 \citealp{Mehdipor2017}, and NGC\,3227; \citealp{Wang2022}). Furthermore, X-ray observations of \mrk\ also showed heavy obscuration \citep{Kara2021, Miller2021}. Changes in X-ray and UV obscuration occur at the same time, suggesting a common origin \citep{Partington2023}.

As for NGC 5548 in STORM 1, the obscuration in \mrk\ seems to have a significant influence on the response of the BLR. In \citet{Kara2021}, hereafter Paper~1, analysis of the first 90 days of the STORM~2 campaign showed decorrelation of the emission-line fluxes from the continuum during the first $\sim55$ days. Analysis of the full campaign by \citealt{Homayouni2023} (hereafter referred to as Paper~2) showed that this was not a persistent decorrelation.  Instead, it was found to occur in multiple temporal windows throughout the campaign, during which the response of the BLR to continuum fluctuations changed dramatically. Contrary to the basic assumption underlying RM analysis, the emission-line light curves are not simply a smoothed, scaled and shifted version of the continuum light curve.

This paper, the fifth in a series describing the AGN STORM~2 results, focuses on the anomalous response of the BLR, particularly for the \ion{C}{4} emission because it is the least contaminated by the obscuring absorption lines, and the least blended with adjacent emission lines. Similar anomalous continuum response has also been reported by \citet{Cackett2023} in studying the Swift light curves. 
The present work has two primary goals. The first is to identify the different temporal windows where the BLR response to the continuum variations significantly changes, and thus affects the measured lag throughout the campaign. The second goal is to understand the role of obscuration in the reprocessing of radiation and its impact on the BLR lag.
We briefly describe the HST observations in \autoref{sec:data}. We present the anomalous BLR variations in \autoref{sec:anomaly}. We discuss the implication of our results in \autoref{sec:discussion} and summarize our findings in \autoref{sec:summary}. We adopt a $\Lambda$CDM cosmology with $\Omega_{\Lambda}$ = 0.7, $\Omega_M$ = 0.3, and $H_0$ = 70 km $\mathrm{s^{-1} Mpc^{-1}}$. Throughout this work, we refer to observation times as ``truncated HJDs" (i.e., THJD = HJD$-$2450000).

\section{The STORM~2 Monitoring Campaign on \mrk}\label{sec:data}
\begin{figure*}[tt]
\centering
\includegraphics[width=0.88\textwidth]{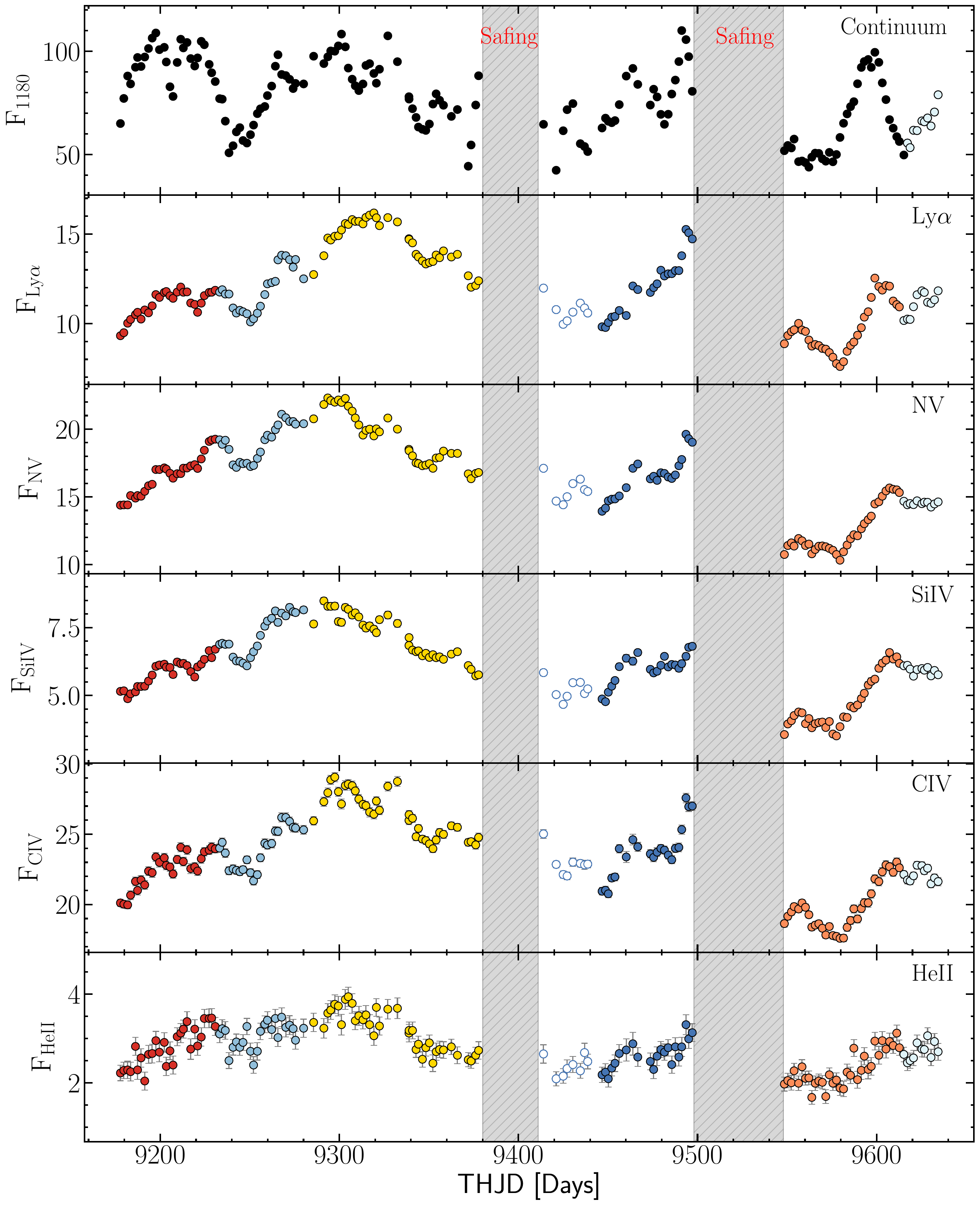}
 \caption{Continuum light curve at 1180 $\mathrm{\AA}$ (top panel) and emission-line light curves for \Lya, \ion{N}{5},  \ion{Si}{4} + \ion{O}{4]}, \ion{C}{4}, and \ion{He}{2}\ (bottom panels). All the light curves are presented in the observed frame. The continuum flux is in units of $10^{-15}~\mathrm{erg\, s^{-1}\, cm^{-2} \AA^{-1}}$ and the line fluxes are in units of $10^{-13}~\mathrm{erg\, s^{-1}\, cm^{-2}}$. The hashed gray regions display the two major observation gaps due to HST safing incidents. The color schemes for the data points correspond to different light curve variation regimes that will be discussed throughout this paper. The open blue symbols between THJD = 9410 -- 9440 (in Window~4) are the eight epochs where the HST observations were more sparsely sampled after an extended safing event and are excluded from the rest of the analysis (see the Appendix for details). The \Lya\ flux is integrated over the blue wing of the emission line to avoid contamination by time-variable absorption features and also the \ion{N}{5} emission. Similarly, the \ion{C}{4} light curve excludes the blue wing of the line profile because of the time-variable absorption (Paper~2). The full \Lya\ and \ion{C}{4} line fluxes are about $5\times$ and $1.6\times$ greater, respectively, than indicated here (see Table 2 of Paper~2). Nevertheless, we present these baseline measurements to demonstrate that, while individual emission line fluxes may not be entirely representative of the total flux, the anomalous characteristics, including those evident in \ion{C}{4}, broadly manifest across all emission lines. 
 }
\label{fig:lc_windows}
\end{figure*}

The AGN STORM~2 program on \mrk\ consists primarily of 165 epochs of HST observations\footnote{HST-GO-16196; \citealt{Peterson2020}} using COS with the G130M and G160M gratings to cover the 1070~\AA\ -- 1750~\AA\ range in single-orbit visits with an approximately 2-day cadence. The HST program began on 2020 November 24 and ended on 2022 February 24. Paper~2 extensively describes the HST program, data products, and COS spectral calibration along with full campaign results \footnote{The AGN STORM~2 data products are available in MAST \url{https://archive.stsci.edu/hlsp/storm2}}. During the UV monitoring, the HST program suffered two extended safe-mode incidents resulting in month-long gaps in the HST UV coverage (see Figure~\ref{fig:lc_windows}). Coordinated photometry and spectroscopy supplemented the HST observations, resulting in full X-ray to near IR coverage of \mrk\ over 15 months (see Paper 1 for a campaign overview). Details of those additional observations will be described in a series of follow-up papers.



\section{Anomaly in the BLR}\label{sec:anomaly}
The year-long monitoring of \mrk\ affords a unique opportunity to study the emission-line variations over an extended period of time. Paper~1 gives an overview of the STORM~2 campaign and its early results. As shown in Paper~1, during the first 90 days of the campaign, the emission-line flux was only weakly correlated with the continuum. Paper~2 shows that the emission-line light curves are not just smoothed, scaled, and shifted transformations of the continuum light curve. Examining light curves for the whole campaign (see Figure~\ref{fig:lc_windows}) shows that even though the continuum at the beginning of the campaign is near its peak brightness, the emission lines are low and rising (red points in Figure~\ref{fig:lc_windows}). 
However, after the emission line peak at THJD~=~9232, the broad line variations became stronger and more representative of the continuum variations. Similar periods of weak correlation reverting to a strong response to 
continuum fluctuations occur throughout the remaining year of the campaign. 

\subsection{One-dimensional Linearized Echo Models}\label{sec:memecho}

\begin{figure*}
\centering
\includegraphics[width=0.98\textwidth]{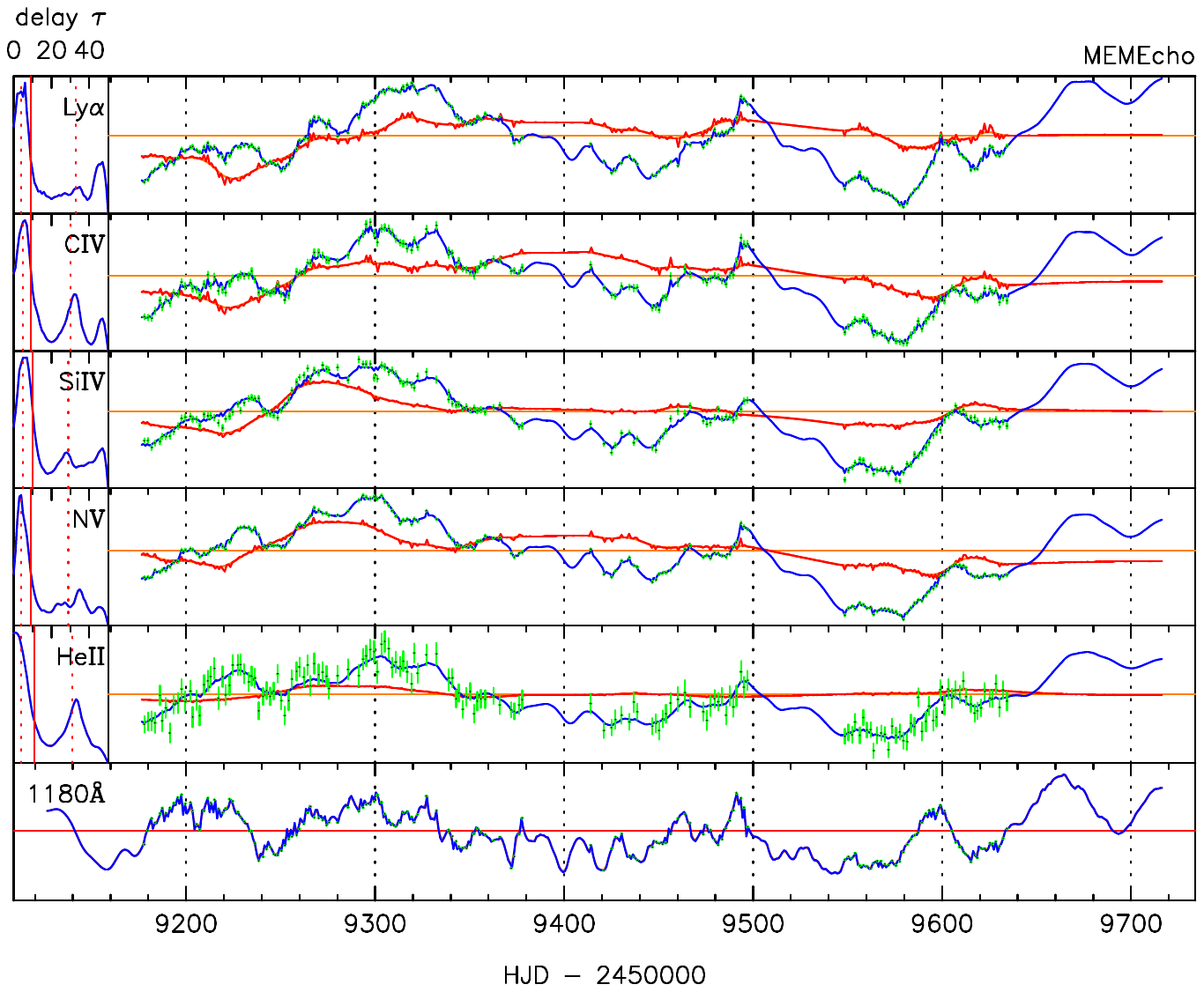}
 \caption{
MEME\textsc{cho} fits to the 1180~\AA\ continuum and five emission-line light curves for Mrk~817 in the STORM 2 campaign. The bottom panel shows the 1180 ~\AA\ continuum light curve (black points with green error bars), which is the driving light curve. The red horizontal line is a reference level at the median of the data. The top panels show the five emission-line light curves (right) and corresponding delay maps (left). The black points with green error bars show the light-curve data, and the blue curves show the echo models. The red curves show the slow background variations. The vertical red line in each delay map marks the median observed-frame delay, and the vertical dashed red lines mark the quartiles of the delay distributions. The MEME\textsc{cho} models account for much of the light-curve structure as echoes of the driving light curve, but they also require significant additional variations (red curves).
 }
 \label{fig:memecho}
\end{figure*}
Inspired by the anomalous emission-line responses exhibited by NGC 5548 during STORM~1, we set out to analyze the light curves for Mrk 817 in STORM 2 in a model-independent way. We used the maximum entropy method (MEM) implemented in a code called MEME\textsc{cho} for estimating time delays in reverberation mapping of AGN \citep{Horne1991, Horne1994} to obtain a one-dimensional linearized echo model \citep{Horne2021}.
Our model uses the 1180 \AA\ continuum light curve, $C(t)$, as the driver, assuming it is a proxy for the ionizing continuum.
For a time delay of $\tau$, the flux $L(t)$ of each emission line is a non-linear function of the continuum light curve shifted to an earlier time, $t - \tau$.
MEME\textsc{cho} linearizes the problem by decomposing the line and continuum light curves into reference levels $L_0$ and $C_0$ with variations $\Delta L(t)$ and $\Delta C(t)$ that are tangent-curve approximations to the parent non-linear functions. With the continuum light curve expressed as

\begin{equation}\label{eq:1}
     C(t) = C_0 + \Delta C(t)
\end{equation}
\noindent
the emission line light curve is then a convolution of the continuum variations
with the one-dimensional delay distribution, $\Psi(\tau)$:

\begin{equation}\label{eq:2}
     L(t) = L_0(t) + \int_ {}^{} \Psi(\tau) \Delta C(t - \tau) d\tau .
\end{equation}

Similar to the analysis of NGC 5548, we allow for a time-dependent background for
each of the modeled emission lines, $L_0(t)$.
The maximum entropy regularization employed by MEME\textsc{cho} keeps the resulting
delay maps positive and produces solutions that are as smooth as possible.

Figure \ref{fig:memecho} shows the results of modeling the STORM 2 light curves with the function in equation \ref{eq:2}.
The driving light curve, $C(t)$, is the 1180~\AA\ continuum shown in the bottom panel,
with the reference level $C_0$ shown as a red horizontal line.
The top five panels show the emission-line light curves with the data points in black,
the error bars in green, and the modelled MEMEecho light curves in blue.
The left column of the figure shows the derived delay maps for each of the emission lines.
The MEME\textsc{cho} solution produces a static, one-dimensional delay map for each line that is a good fit to the data. However, it does require large variations in the background levels for each emission line, $L_0(t)$, as shown by the red curves in each panel.

Significantly, the large excursions below the mean from the beginning of the campaign to
THJD$\sim$9260, and again around THJD=9600 correspond to temporal windows during the
STORM 2 campaign when absorption was the strongest (as we will show later).
Similarly, absorption was weakest during the time interval from THJD$\sim$9300--9400
when the background levels vary above the mean.

The delay maps for each emission line have strong peaks at delays of $\sim$5 days, with \ion{He}{2}
showing the shortest delay. While the central region is where most of the response is
located, there is a secondary peak at delays of $\sim$30 days, particularly for
\ion{C}{4} and \ion{He}{2}.

\subsection{Understanding the Anomalous \CIV\ Light Curve}\label{sec:lag_windows}
\begin{figure*}
\centering
\includegraphics[width=0.98\textwidth]{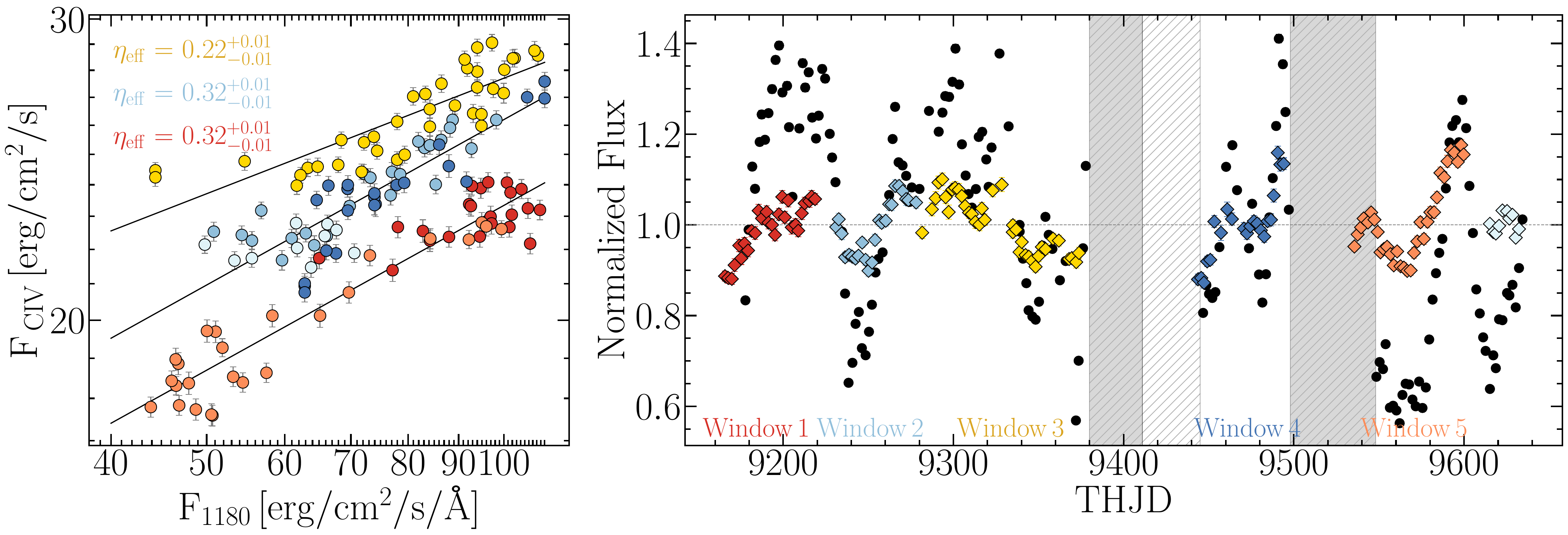}
 \caption{Left: The ``time-delay corrected" $\rm F_{\rm CIV}$ (in $10^{-13}~\mathrm{erg\, s^{-1}\, cm^{-2} \AA^{-1}}$) vs. the $\rm F_{1180}$ continuum (in $10^{-15}~\mathrm{erg\, s^{-1}\, cm^{-2} \AA^{-1}}$) measurements displayed on a logarithmic scale. Here the $\rm F_{\rm CIV}$ is ``time-delay" corrected based on lag measurements in Table~\ref{tab:table2} (see Section \ref{sec:c4_lags} for detail). The color coding represents the identified windows in Table~\ref{tab:table1} using the same color scheme as in Figure~\ref{fig:lc_windows}. We identify three main trends in the \CIV\ emission-line response to continuum variations. The best-fit slope to each trend is illustrated with a black line representing the emission-line response, $\eta_{\rm eff}$, for each group of data points. In contrast to the majority of points which show a positive correlation between $\rm F_{CIV}$ and $\rm  F_{1180}$, the observations during the last $\sim$3 weeks of data (light blue points) show no correlated variability, and have a flat response. Right: The \ion{C}{4} light curve (colored symbols) superposed on the continuum light curve (black symbols). The continuum light curve is normalized to its median; the \ion{C}{4} light curve is normalized to the median in each segment, and each segment is shifted based on the lag measurement in each of the identified windows in Table~\ref{tab:table1}. We adopt a zero time delay for the last $\sim$3~weeks of data (light blue data points) as the lag measurement in the final window is consistent with a zero lag.
 }
\label{fig:flux_flux}
\end{figure*}

Although the MEME\textsc{cho} analysis successfully yields static one-dimensional delay maps for the BLR, there is no inherent physical motivation accounting for the slowly varying background required by the model. However, these are real, significant variations the physics of which are not well understood and the MEME\textsc{cho} approach is only driven by the data behavior. Here we develop a plausible physical explanation for the time-varying background. We start with the anticipated correlation between the flux in the \ion{C}{4} emission line, $F_{\rm C\,IV}$, and the continuum flux, $F_{\rm 1180}$. We choose $F_{\rm 1180}$ as our reference since it is the closest uncontaminated continuum window (by absorption) to the ionizing continuum. As shown in Figure~\ref{fig:flux_flux}, while there is an overall positive correlation between the delay-shifted $F_{\rm C\,IV}$ and $F_{\rm 1180}$, the $F_{\rm C\,IV}$ rises by 25\% while $F_{\rm 1180}$ doubles (see Section~\ref{sec:varying_response}), there is $8$\% root-mean-square (RMS) scatter around the best-fit line to the full campaign. The $F_{\rm C\,IV}$ - $F_{\rm 1180}$ correlation seem to show a similar marginal response (see $\eta_{\rm eff}$ reported in Figure~\ref{fig:flux_flux}) throughout the campaign.
However, if we examine this correlation with points selected by observation time (THJD), we see that for several temporal windows, the normalization appear to change between one temporal window to the next.
In fact, we can measure an independent and slightly different $F_{\rm C\,IV}\propto F_{1180}$ response relation for each of these temporal windows, which will reduce the RMS scatter around each individual best-fit line to $\approx2-3\%$. This may be an indication of a reduction in ionizing photons incident on BLR gas, not tracked by continuum luminosity changes. 
Table~\ref{tab:table1} summarizes the start and end times of these temporal windows (details of the selection process for the window boundaries is given in the Appendix).
Each window is indicated by a different color in Figure \ref{fig:flux_flux}, with corresponding power-law response relations for different families of points overlayed on the plot (see the left panel in Figure \ref{fig:flux_flux}). 

The temporal windows corresponding to these families of points are color coded similarly to the light curves shown in Figure \ref{fig:lc_windows}. All the UV emission lines except \ion{C}{4} are affected by 
broad absorption troughs or blended emission line wings (see Paper~1 and
Paper~2), and thus require detailed modeling and de-blending (Kriss et~al. in prep). 
Thus, for the remainder of this work, we concentrate on the \ion{C}{4} light curve behavior.
Below, we discuss in more detail the time-delay measurements for \ion{C}{4}
as a function of the significant variations with time in its response to the
observed continuum.

\begin{table*}[t]
\caption{Emission Line Response Windows}
\begin{center}
  \begin{tabular}{cccc}
    \hline
    \hline
    Window & THJD & Calendar Date & Duration\\
    {} & {Days} & {Days} & {Days} \\
    \hline
    1  & 9177 to 9232 & 2020-11-24 to 2021-01-18 & 55 \\
    2  & 9232 to 9282 & 2021-01-18 to 2021-03-09 & 50 \\
    3  & 9282 to 9378 & 2021-03-09 to 2021-06-13 & 96 \\
    Gap & 9378 to 9413 & 2021-06-13 to 2021-07-18 & 36 \\
    4$^{\dagger}$  & 9413 to 9498 & 2021-07-18 to 2021-10-11 & 87\\
    Gap & 9498 to 9548 & 2021-10-11 to 2021-11-30 & 51 \\
    5  & 9548 to 9615 & 2021-11-30 to 2022-02-05 & 67 \\
    6  & 9615 to 9634 & 2022-02-05 to 2022-02-24 & 19\\
    \hline
 \end{tabular}
  \tablecomments{
\footnotesize
${}^{\dagger}$ We later redefine Window~4 in Table~\ref{tab:table2} to include only THJDs from 9445 to 9498. Also, see the discussion in the Appendix~\ref{sec:appendix}.
}
 \label{tab:table1}
\end{center}
\end{table*}
 
\subsection{Time-varying Lags for \CIV\ }\label{sec:c4_lags}
Previous ground-based optical RM campaigns targetting \mrk\ \citep{Peterson1998, Denney2010} successfully measured \Hb\ time-delays of 14--34 days relative to the 5100~\AA\ continuum. Early campaign results in Paper 1 suggest that there is a $\sim$ 4-day time delay between the $\rm F_{1180}$ and Swift V-band. Paper~1 Figure~14 shows a \Hb\ lag of 23.2 $\pm$ 1.6 days behind the continuum. 
The $\CIV$ time-delay is expected to be half of the observed \Hb\ lag, motivated by \citet{Lira2018}.
Paper~2 finds a time-delay of $11.8_{-2.8}^{+3.0}$ days between the $F_{1180}$ and the \CIV\ emission line using the full set of HST UV observations, which is consistent with this prediction. However, as we argued above, this single time-delay measurement does not fully capture the diverse line responses.
Therefore, here we study each of the identified windows in
Table~\ref{tab:table2} independently.

\begin{figure}
\centering
\includegraphics[width=0.28\textwidth]{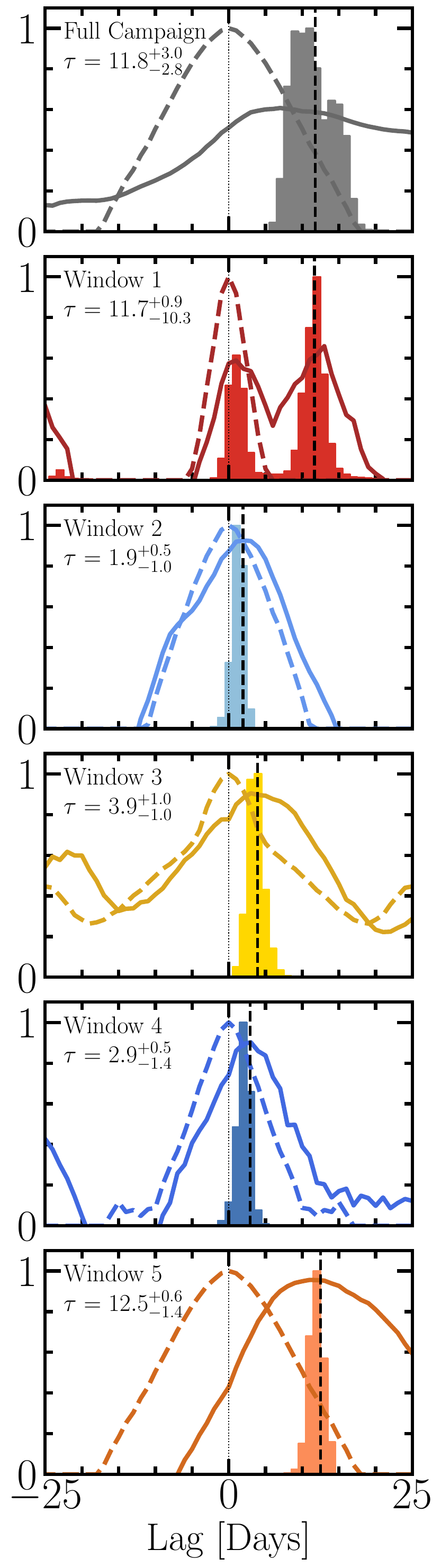}
 \caption{The cross-correlation function (CCF; solid line) between $F_{1180}$ continuum and \ion{C}{4} in the full campaign (top panel) compared with the CCF computed for each time interval defined in Table~\ref{tab:table2} (bottom panels). The dashed line shows the auto-correlation function (ACF) of the continuum in respective time intervals along with the \pyccf\ cross-correlation centroid distribution (CCCD). The rest-frame lag for each time interval is shown by a vertical dashed line and also reported in Table~\ref{tab:table2}. We find that the \ion{C}{4} CCCD changes significantly over each of the windows, where Windows~1 and 5 prefer a long lag ($\approx$12~days), while Windows~2, 3, and 4 have a shorter lag ($\approx$2-4 days). Window~1 shows a secondary peak at a shorter time delay that is similar to Windows~2, 3, and 4. However, the longer lag seems to be the dominant peak in Window~1 CCCD.}
 \label{fig:bimodal_lag}
\end{figure}

\begin{table}[t]
\caption{\CIV\ Lag Measurement (\pyccf)}\label{tab:table2}
\begin{center}
  \begin{tabular}{ccccc}
    \hline
    \hline
    Window & THJD & Time Delay & $r_{\rm max}$ & N Data Points\\
    {} & {Days} & {Days} & {} \\
    \hline
    1  & 9177$-$9232 &  $11.7_{-10.3}^{+0.9}$ & 0.66 & 28 \\
    2  & 9232$-$9282 & $1.9_{-1.0}^{+0.5}$ & 0.93 & 24\\
    3  & 9282$-$9378 & $3.9_{-1.1}^{+1.0}$ & 0.91 & 38\\
    4$^{\dagger}$ & 9445$-$9498 & $2.9_{-1.4}^{+0.6}$ & 0.90  & 22\\
    5  & 9548$-$9615 & $12.5_{-1.3}^{+0.6}$ & 0.96 & 34\\
    \hline
 \end{tabular}
 \tablecomments{
\footnotesize
${}^{\dagger}$ Window~4 is shortened to remove the sparse sampling interval of $\sim$ 30 days that immediately occur after the first safing event (see Appendix~\ref{sec:appendix} for a detailed discussion).
}
\end{center}
\end{table}
We adopt the python implementation of the commonly-used time-series analysis method CCF (\pyccf; \citealp{Sun2018b})
to measure the mean time delay between the $\rm F_{1180}$ continuum and the \CIV\ emission-line flux variations and compute the cross-correlation Pearson coefficient $r$ as a function of time delay $\tau$ (often referred to as the Interpolated Cross-Correlation Function, ICCF \citealt{Gaskell1986, Gaskell1987, White1994}) for each of the five temporal windows.
We use $\pm$25 days for the lag search range in each of the temporal windows, though we use $\pm$50~days for the full campaign results (top panel in Figure~\ref{fig:bimodal_lag}). We estimate the uncertainty in $\tau_{\rm ICCF}$ using Monte Carlo simulations that employ the flux randomization and random subset sampling (FR/RSS; \citealp{Peterson1998}). We adopt 20000 Monte Carlo (MC) iterations to obtain the cross-correlation centroid distributions (CCCDs, \citealp{Peterson1998, Peterson2004}). To ensure that the time-delay measurements are not due to our choice of lag measurement method, we also use the Z-transformed discrete cross-correlation function (ZDCF) approach \citep{Alexander1997, Alexander2013, Kovacevic2017} in combination with a Gaussian process regression (GP) to model the stochastic AGN light curves with arbitrary sampling. We find that the ZDCF approach recovers similar time-delay measurements as the PyCCF approach in each of the temporal windows, with time delays consistent within 1$\sigma$. We report rest-frame lag measurements in Table \ref{tab:table2}, along with cross-correlation coefficients, and number of datapoints in each of the temporal windows. The CCCDs from PyCCF reported in Figure~\ref{fig:bimodal_lag} reveals significantly different results for each of the temporal windows. The CCCDs shows a clear bimodal distribution across the five temporal windows where two typical time lag results are measured: one at 2-3 days and one at 11-12 days. In particular, the mean time delays corresponding to Windows~1 and 5 (i.e., red and orange in Figure~\ref{fig:flux_flux}) are significantly
longer than the time delay measured for Window~3 (gold), which is slightly longer than ones obtained for Windows~2 and 4 (light and dark blue). The cross-correlation coefficient distribution in Window~1 
shows a secondary peak that coincides with the short time delays of Windows~2, 3, and 4. Also, the FR/RSS uncertainty range for Window~1 encompasses lags consistent with 1~day, and the $r_{\rm max}$ is lower compared to temporal windows~2$-$5. Despite these anomalous features in the Window~1 CCCD, the peak at $\sim$11 days has a higher maximum cross-correlation coefficient $r_{\rm max}$, and thus we adopt this as the primary peak during Window~1. Although our chosen windows have sharp boundaries, we have measured the lag in the time intervals both assuming a sudden transition between the intervals as well as a smoother $\sim$10-day transition for each time interval and find that the measured lags are consistent to within 1$\sigma$. 

\begin{figure*}
\centering
\includegraphics[width=\textwidth]{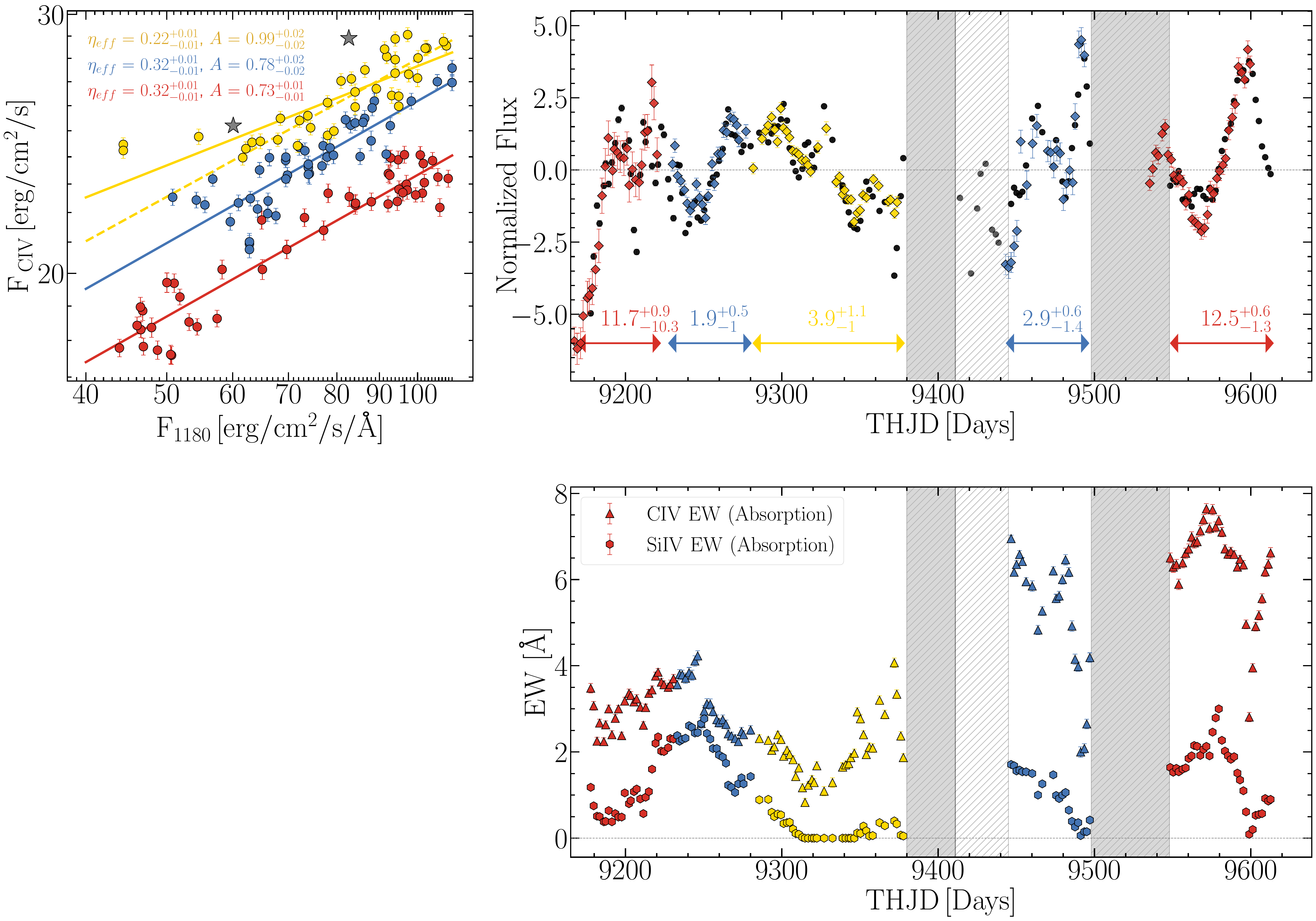}
 \caption{Upper Left: Time-delay-corrected fluxes for $ F_{\CIV}$ (in $10^{-13}~\mathrm{erg\, s^{-1}\, cm^{-2} \AA^{-1}}$) vs. the $F_{1180}$  continuum (in $10^{-15}~\mathrm{erg\, s^{-1}\, cm^{-2} \AA^{-1}}$) plotted on a logarithmic scale (similar to Figure~\ref{fig:flux_flux} but with a different color coding). The color coding represents the three main trends in the \CIV\ emission-line response to continuum variations in Table~\ref{tab:table3}. The best-fit slope to each trend is illustrated with a colored line representing the emission-line responsivity, $\eta_{\rm eff}$ and $A$ is the best-fit line intercept, as given in the upper left corner (see equation~\ref{eq:3}). The grey stars from archival measurements (uncorrected for time-delay) of Mrk~817 obtained in 2009 suggest a significantly higher response and EWs at that time. The yellow dashed line is the fit excluding the three outlier data points in the gold data points.  
 Upper Right: The continuum flux $F_{1180}$ (black symbols) overplotted on the $F_{\CIV}$ light curve (colored symbols) where each light curve is normalized to a median of zero and an normalized median absolute deviation (NMAD) of one and also corrected for the mean time delay in each separate window. Major gaps due to HST safing events are identified by the gray shaded regions. The \CIV\ light curve exhibits a varied behavior in response to the continuum, with the red light curve segments showing a lag of $\sim$ 11-12.5 days, the blue segments a lag of $\sim$ 2.5 days, and the gold segment a lag of $\sim$ 4~days. Lower Right: Time series of the absorption as indicated by \ion{Si}{4} and \ion{C}{4}. Each interval is color-coded by the same colors as the light curve. The horizontal line shows the absorption level at zero. The gold symbols corresponds to the window where absorption is at its minimum.
 }
\label{fig:lc_overlap}
\end{figure*}

As a visual verification of the measured time delays, the right panel of Figure~\ref{fig:lc_overlap} presents the overlapping continuum and ``delay-corrected" \CIV\ light curves, which are color coded by the three identified main trends. For illustrative purposes, in each window, the continuum and \CIV\ light curves are normalized to a median of zero and a normalized median absolute deviation (NMAD e.g., \citealt{Maronna2006}) of unity. This preserves the shape and variability amplitude of the light curves while enabling a one-to-one comparison between the continuum and the emission line light curve features. In general, the normalized and shifted \CIV\ light curves using the individual-window lags provide a reasonable match to the features observed in the continuum.
The overlapping light curves are also especially instructive for segments of the temporal windows that do not overlap with the continuum features and require careful consideration.

\subsection{Variations in the \ion{C}{4} Response} \label{sec:varying_response}
To study the time-dependent variations in the \ion{C}{4} emission-line
response to the continuum fluctuations, we follow \citet{Goad2016} in
connecting the 1180 \AA\ continuum and \CIV\ reprocessing
efficiency $\eta_{\rm eff}$ using the
$F_{\rm \CIV} \propto F_{1180}^{\eta_{\rm eff}}$ correlation.
We treat the temporal windows identified in Table~\ref{tab:table2} as independent segments of the light curve and shift back in time each segment of the emission-line light curve by its respective time delay.
For the time-delay-corrected continuum flux, we select the continuum
flux point closest in time to the shifted emission-line flux.
We emphasize that because of the existence of two extended gaps
in our observations and the complex light curve behavior, implementing the weighted approach of \citet{Goad2016} was not feasible for reconstructing the continuum flux.
Furthermore, due to the presence of gaps or data associated with the beginning of the campaign, a subset of delay-corrected emission line fluxes is associated
with the same continuum flux measurements, and thus form a cluster of
overlapping points in Figure~\ref{fig:flux_flux}.
This is more evident in windows~1 and 5.
We include only the first overlapping entry for these clusters of
points and exclude the rest from our analysis.
We combine all the delay-corrected continuum and emission line fluxes and
obtain the time-averaged emission line response $\eta_{\rm eff}$ using
the relation
\begin{equation}\label{eq:3}
    \mathrm{log} \,F_\mathrm{CIV}  = A + \eta_{\rm eff}\,\mathrm{log}\, F_{1180}
\end{equation}
Where $A$ is related to the characteristic EW of \ion{C}{4} at $\rm F_{1180} = 75\times 10^{-15}\, erg\,s^{-1}\,cm^{-2}\, \AA^{-1}$, $A = \rm log[75^{(1-\eta)}\times EW_{CIV}{ (@75\times 10^{-15})/100\,\, \AA}]$.
We use a linear regression method including uncertainties in $F_{\mathrm{CIV}}$ and $F_{1180}$ to determine the \ion{C}{4} marginal response and reprocessing efficiency as identified by best-fit slope $\eta_{\rm eff}$ and normalization $ A$. We adopt the \texttt{SciKitLearn} linear regression model to perform an ordinary least squares linear regression.
We report the best-fit value and 1$\sigma$ uncertainties using bootstrap sampling of light curves with 1000 realizations with replacement. We report the best-fit $\eta_{\rm eff}$ and $ A$ in Table~\ref{tab:table3}.
We combine groups of points with similar \ion{C}{4} flux response at a given continuum flux. 
We identify three main trends based on the value of the \ion{C}{4} flux at that continuum level, corresponding to Windows~1 and 5, Windows~2 and 4, and Window~3. These relations are shown as best-fit lines in the left panel of Figure~\ref{fig:lc_overlap}, with gold corresponding to the high reprocessing efficiency in Window~3 (higher \ion{C}{4} flux at fixed continuum), red the low reprocessing efficiency in windows~1 and 5 (lowest \ion{C}{4} flux at fixed continuum), and blue the intermediate reprocessing efficiency in windows~2 and 4. We also show these three relations as colored lines in Figure \ref{fig:lc_overlap} and Table~\ref{tab:table3}.

The three lowest $F_{1180}$ flux points in Window~3 seem to influence the slope of the best-fit line. When these three points are removed, the slope is similar to the two lower trends, $\eta_{\rm eff} = 0.31\pm0.01$. However, these three points correspond to the three data points immediately before the first safing event, so it is plausible that the delay-corrected emission line fluxes immediately before the safing gap may not be reliable. This may be because the light curve is already changing but due to overlap with the safing gap it is not captured in the data.
The best-fit line to Window~3 with and without the three outlier points is shown in the left panel of Figure~\ref{fig:lc_overlap}. The best-fit values are also reported in Table~\ref{tab:table3}. While removing the three points changes the Window~3 slope (dashed yellow line) to be similar to the two lower trends, the intercept remains significantly larger than the two other trends. This suggests that there may be a difference in the underlying emission line flux distribution between Window~3 and the other two windows. While it is plausible that the total amount of reprocessing might have changed between the temporal windows since the reprocessing efficiency changes per fixed continuum flux level, the slope similarity suggests that the effective responsivity is approximately the same in each window. This could be caused by a change in the fraction of ionizing photons intercepted by BLR gas (i.e., obscuration) or changes in the spectral shape, in which the extreme ultraviolet (EUV) flux has changed relative to the flux of the proxy continuum $F_{1180}$ between one temporal window to the next.

\begin{table*}[t]
\caption{\CIV\ Response Measurements}\label{tab:table3}
\begin{center}
  \begin{tabular}{cccccc}
    \hline
    \hline
    Window & THJD & $\eta_{\rm eff}$ & $A$ & Characteristic \ion{C}{4} EW & Best-fit RMS  \\
    {} & {Days} & {} & {} & {\AA} & {} \\
    \hline
1 & 9177 to 9232 & $0.16_{-0.03}^{+0.02}$ & $1.05_{-0.05}^{+0.05}$ & $29.9\pm0.7$ & 2.65\% \\ 
2 & 9232 to 9282 & $0.29_{-0.01}^{+0.01}$ & $0.84_{-0.02}^{+0.03}$ & $32.3\pm0.6$ & 2.55\% \\ 
3 & 9282 to 9378 & $0.22_{-0.01}^{+0.01}$ & $0.99_{-0.02}^{+0.02}$ & $33.7 \pm 0.9$ & 3.17\% \\ 
$3^*$ & 9282 to 9378 & $0.31_{-0.01}^{+0.01}$ & $0.82_{-0.02}^{+0.02}$ & $33.6 \pm 0.9$ & 3.12\% \\ 
4 & 9445 to 9498 & $0.38_{-0.02}^{+0.01}$ & $0.65_{-0.03}^{+0.03}$ & $30.7\pm 0.8$ & 3.26\%\\ 
5 & 9548 to 9615 & $0.31_{-0.01}^{+0.01}$ & $0.74_{-0.02}^{+0.02}$ & $27.9\pm 0.6$ & 3.12\%\\ 
6 & 9615 to 9634 & $0.06_{-0.04}^{+0.04}$ & $1.24_{-0.06}^{+0.06}$ & $30.0\pm 0.5$ & 1.89 \%\\   
1 + 5 & 9177 to 9232 and 9548 to 9615 & $0.32_{-0.01}^{+0.01}$ & $0.73_{-0.01}^{+0.01}$ & $28.5\pm 0.7$ & 3.28\% \\
2 + 4 & 9232 to 9282 and 9445 to 9498 & $0.32_{-0.01}^{+0.01}$ & $0.78_{-0.02}^{+0.02}$ & $32.0\pm 0.8$ &3.24\%\\
 
 \hline
 \end{tabular}
 \tablecomments{
\footnotesize
${}^{*}$ We exclude the three data points with lowest flux in Window~3 and report the fit values.
\newline
The characteristic \ion{C}{4} EW is the ratio of line flux measured when $\rm F_{1180} = 75 \times 10^{-15}\, erg\,s^{-1}\,cm^{-2}\, \AA^{-1}$ to the continuum $\rm F_{1180} = 75 \times 10^{-15}\, erg\,s^{-1}\,cm^{-2}\, \AA^{-1}$. We note that the \ion{C}{4} EW was measured using the red wing of the \ion{C}{4} profile due to contamination with absorption. The EW is $\approx$60\% of the estimated total flux in \ion{C}{4}, as reported in Paper~2.
}
\end{center}
\end{table*}

\subsection{Obscuration and the \CIV\ Response}
A significant element affecting the BLR response in Mrk 817 during STORM2 is the
presence of outflowing gas that obscures the X-ray and ionizing continuum (Paper~1).
To establish an unobscured baseline, we examine archival spectra of \mrk\, when it was observed in 2009 using the HST COS instrument \citep{Winter2011}.
During these archival observations, the UV spectrum showed no broad absorption
troughs (See Figure~2 of Paper~1 for a comparison between the 2009 archival
spectra and the AGN STORM~2 campaign). We use the archival spectra to measure the continuum and \ion{C}{4} flux, adopting the continuum windows 1493$-$1511 $\rm \AA$ and 1736$-$1741.5 $\rm \AA$ and emission line integration limit 1590$-$1638 $\rm \AA$ as in Paper~2. 
We find that the unobscured \ion{C}{4} flux response was significantly higher in 2009 (see the gray star symbols in Figure~\ref{fig:lc_overlap}). While these \ion{C}{4} fluxes are higher, and the slope connecting them is steeper than the trends observed in our campaign in $F_{1180}-F_{\CIV}$ relation, this cannot be verified since these two isolated archival spectra measurements cannot be placed in context without contemporaneous continuum monitoring to obtain a lag and perform a time-delay correction.
 
\begin{figure*}
\centering
\includegraphics[width=\textwidth]{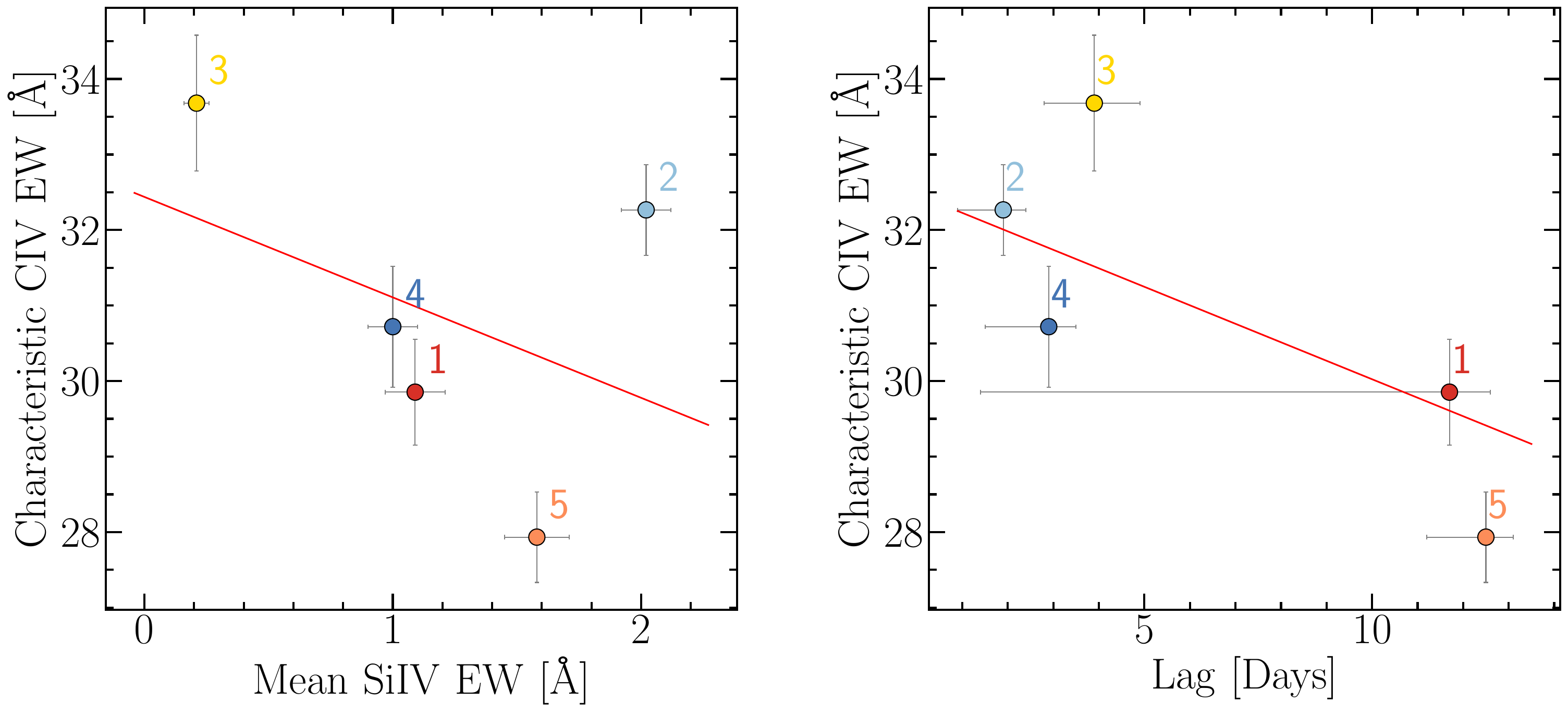}
 \caption{Left: Characteristic \ion{C}{4} EW (as given in Table \ref{tab:table3}) versus the mean equivalent width of broad \ion{Si}{4} absorption as a proxy for obscuration in the corresponding time interval. We find smaller characteristic \ion{C}{4} EW during the temporal windows where the BLR is heavily shielded from the ionizing continuum. Right: Characteristic \ion{C}{4} EW versus the time lag in the corresponding time interval as given in Table \ref{tab:table2}. We find longer time delays during the temporal windows where line response is the lowest (see Figure~\ref{fig:blr_cartoon}).
 }
\label{fig:correlations}
\end{figure*}

The appearance of the intrinsic UV absorption and blending of the emission lines during the AGN STORM~2 campaign complicates the analysis of the individual spectra. Even though the \ion{C}{4} emission line is the least affected emission line, to properly investigate a possible phenomenological connection between the intrinsic absorption and the complex \ion{C}{4} response, we use a heuristic spectral model to disentangle these complications in the individual spectra. Our spectral modeling is described in Paper 1, and it follows the approach adopted by \citet{Kaastra2014} and \citet{Kriss2019b} for NGC\,5548. We use a reddened power-law continuum plus multiple Gaussian components to qualitatively fit the emission and broad absorption features. We use multiple Gaussian components to model the emission lines, and we also measure the variable intrinsic broad absorption features, focusing on the strongest ones (\ion{P}{5}, \ion{C}{3}*, Ly$\alpha$, \ion{N}{5}, \ion{Si}{4}, and \ion{C}{4}). We measure the equivalent widths (EWs) of each absorption line in the normalized spectrum by integrating over pixels lying in the wavelength window of an absorption line. We use this EW as a measure of the strength of the UV absorption and obscuration throughout the campaign. For illustrative purposes, in this paper we focus on the broad \ion{Si}{4} absorption since it is a well resolved doublet, though we note that all the other broad absorption lines behave similarly. It is important to note here that the absorption EW is measured along our line of sight, but it is not clear how well it tracks the average EW of the absorption present over all lines of sight between the continuum and BLR at a given time.

The \ion{C}{4} line flux in any given window depends on the overall flux from the ionizing continuum that reaches the BLR, and the shape of the transmitted ionizing flux in the SED. As one can see from the large scatter in the upper left panel of Figure \ref{fig:lc_overlap}, the observed UV flux $F_{1180}$ appears to be a poor proxy for the ionizing flux, likely due to the strong and variable obscuration. The overall ionizing flux is governed by both the covering factor and column density of the obscurer, while the column density of the obscurer largely governs the shape of the transmitted flux.
The characteristic EW as shown in Table~\ref{tab:table3} and Figure~\ref{fig:correlations} reflects the strength of the reprocessing efficiency for a given time interval and is governed by the total ionizing flux relative to the observed UV continuum. The marginal response of the emission line, $\eta_{\rm eff}$, is determined by the shape of the SED in the ionizing UV. In other words, the \ion{C}{4} line flux is a function of the amount of ionizing continuum that reaches the BLR and the shape of the ionizing continuum. The characteristic EW reflects how much of the ionizing continuum is reprocessed into \ion{C}{4} emission while the marginal response reflects how the \ion{C}{4} emission responds to changes in the ionizing continuum.

The bottom right panel in Figure~\ref{fig:lc_overlap} shows the \ion{C}{4} and \ion{Si}{4} absorption light curves. Comparison of the variation in the $F_{1180}$ continuum and \ion{C}{4} emission-line with the \ion{C}{4} (top panel of \ref{fig:lc_overlap}) and the \ion{Si}{4} absorption light curves (bottom panel of \ref{fig:lc_overlap}) shows that the temporal windows with the strongest absorption correspond to times when the \CIV\ EW is smallest (i.e., the reprocessing efficiency is the smallest). Also, but less significantly, weak absorption corresponds to shorter time lags. Figure \ref{fig:correlations} illustrates these trends more quantitatively. Both panels in Figure \ref{fig:correlations} show how the normalization of the \CIV\ response function is related to the strength of the broad \ion{Si}{4} absorption (left), and to the measured time lag for that time interval (right).
For \ion{Si}{4} the Pearson correlation coefficient $r=-0.43$ with $p = 0.47$, and for the relation versus lag, $r=-0.80$ and $p = 0.10$.
Neither trend is statistically significant, primarily due to the large uncertainties, but the
qualitative sense one obtains in comparing the light curves is borne out by the trends in the scatter plots, where stronger \CIV\ emission-line flux and EW corresponds to weaker absorption and shorter time lags. This trend is also 
consistent with the MEMEcho results (Section~\ref{sec:memecho}). Namely, that the difference in the characteristic EW of \ion{C}{4} as seen in Table~\ref{tab:table3}, the top left panel of Figure~\ref{fig:lc_overlap}, and both panels of Figure \ref{fig:correlations} is generally consistent with the time-dependent background found by MEMEcho, i.e., MEMEcho requires a decrease in the relative strength of the \ion{C}{4} line flux, as measured by its characteristic EW, in temporal windows~1 and 5 (around THJD 9200 and 9600 days). This is in contrast to temporal window~3, where the \ion{C}{4} flux increases, as measured by the larger characteristic EW. 

\begin{figure*}
\centering
\includegraphics[width=\textwidth]{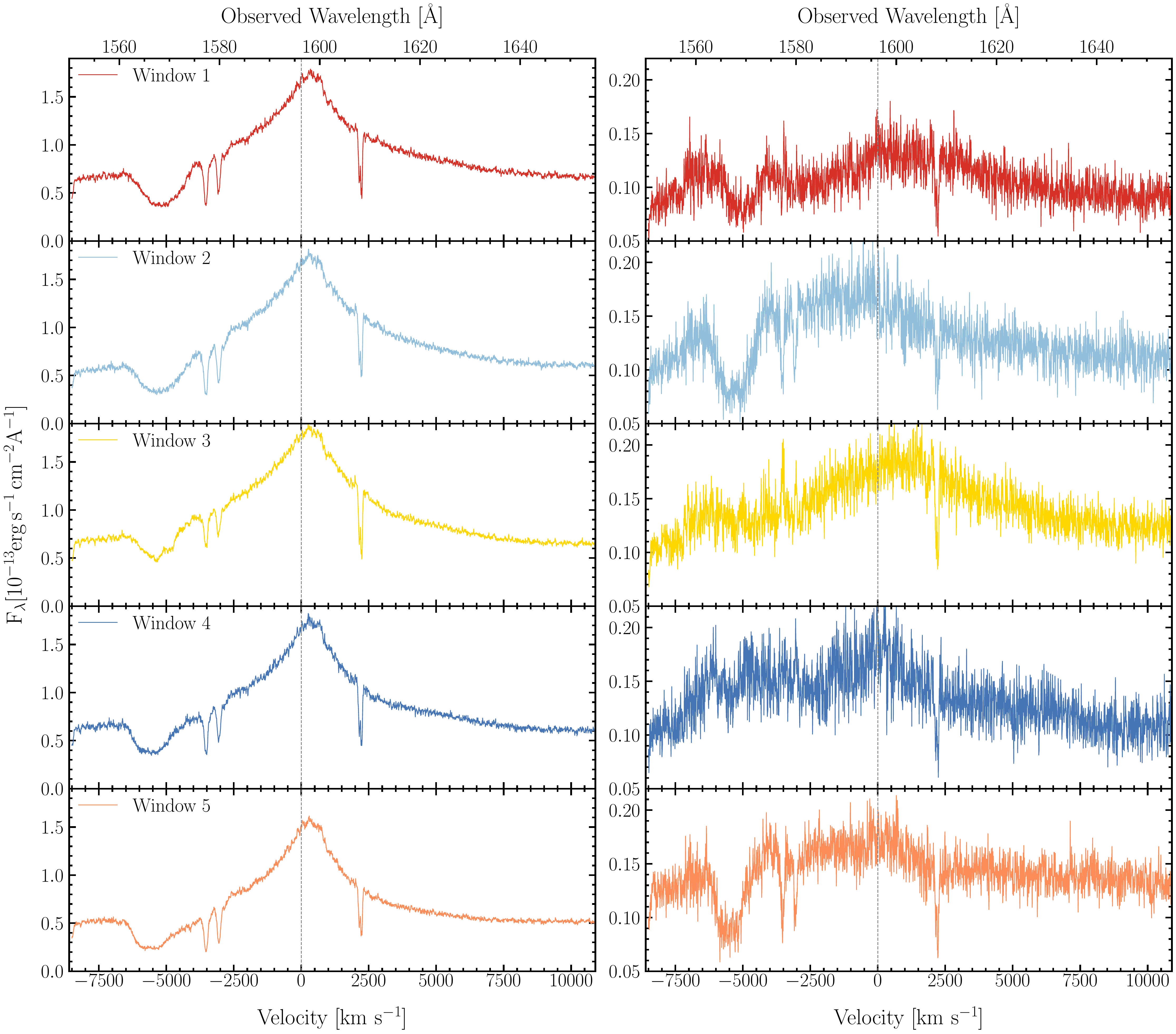}
\caption{The mean (left) and RMS (right) spectra for the temporal windows in Table~\ref{tab:table3}. The RMS profile shows significant variations from one temporal window to the next. Other than the primary peak at zero velocities in the RMS profile, Window~2 (light blue) shows the appearance of a secondary peak in negative velocities that significantly varies in the subsequent temporal windows. The location of this variation coincides with the broad absorption trough. Also, Window~1 shows the smallest variation amplitude in the RMS profile, while Window~2 shows a significantly higher variation amplitude.}
\label{fig:mean_rms}
\end{figure*}

\subsection{Mean and RMS Spectra}\label{sec:binned}
To recover any signature of kinematic information about the BLR and the implications of the light curve variation, we perform a preliminary comparison of the mean and RMS spectra in each of the temporal windows. Similar to Paper~2, we isolate the \ion{C}{4} emission line (see Section~4.2 in Paper~2).
Figure~\ref{fig:mean_rms} shows the mean and RMS spectra for the temporal windows in Table~\ref{tab:table3}. The RMS profile, which contains information about the variable part of the spectrum, shows that the most responsive portion of the \ion{C}{4} profile is in the blue side of the profile at negative velocities, and is changing significantly from one temporal window to the next. This may be due to the outflowing wind into our line of sight. Other than the primary peak near the line center at zero velocities that is present in all the temporal windows, Window~2 and Window~4 (light and dark blue) show the appearance of a secondary peak at negative velocities. This secondary peak significantly varies in the subsequent temporal windows, and the location of this variation coincides with the broad absorption trough. Also, Window~1 shows the smallest variation amplitude in the RMS profile, while Window~2 shows a significantly higher variation amplitude. Initial velocity-binned results from these temporal windows indicate that the signal-to-noise ratio (SNR) is not uniform across all five temporal windows. Consequently, it is challenging to accurately recover information about the dynamics of the \ion{C}{4} emitting region. Therefore, we will defer the two-dimensional RM analysis similar to that of \citet{Horne2021} until we have completed the modeling and corrected for absorption contamination.

\section{Discussion}\label{sec:discussion}

Paper~2 concluded that the emission line light curves cannot be explained by a single, static response for the ultraviolet broad lines that persists for the duration of the campaign. They are not simply a delayed, 
proportional response to the continuum variations.
Above we showed that by examining individual time segments of the
emission-line light curves, we can identify temporal windows where the
emission-line gas responds coherently to the continuum variations.
Using these temporal windows, we show that
the emission-line gas is responding to continuum variations, but the \ion{C}{4} reprocessing efficiency is different from one temporal window to the next.
We now discuss possible physical explanations for these variations in
emission-line response.


Our hypothesis is that as a consequence of the changes in the ionizing flux illuminating the BLR, due to the evolving properties of the obscuring wind, the region of the BLR responding to the continuum fluctuations changes, leading to changes in the measured time delays. However, the time delays measured in the separate temporal windows behave in a counterintuitive way. During periods of heavy obscuration when the ionizing flux is presumably suppressed, one might expect the time lag to be shorter as the ionization zone producing the maximum \ion{C}{4} flux would be closer to the active nucleus, an effect often described as a “breathing mode” \citep{Gilbert2003, Korista2004, Cackett2006, Denney2009, Park2012, Barth2015, Runco2016}. In contrast, we find longer lags when the line-of-sight obscuration is stronger, and shorter lags when it is weaker.

The broad extent of the CCF in all the windows (and in all the lines; see the top panel in Figure~\ref{fig:bimodal_lag}) and from the range in lags from the UV to \Hb\ (Paper~1) shows that the BLR spans radii more than an order of magnitude in size. Describing such a broad distribution with a single number can be misleading since the geometrical distribution of the BLR is convolved with an equally complicated power spectrum of time-variable illuminating radiation. \cite{Goad2014} studied the consequences of photoionizing a BLR with a large range in radial extent and how it relates to the timescale of the continuum fluctuations. Their simulations show that for rapid continuum fluctuations that occur on timescales shorter than the light-crossing time of the BLR, there is a significant dilution of the observed response of the emission line due to two geometric factors: its finite emissivity volume and its non-negligible thickness. Firstly, the BLR is not a point source but possesses a characteristic size, often quantified by an emissivity-weighted radius for each emission line. Secondly, the BLR is not a thin, two-dimensional shell but rather a spatially extensive region with inherent depth.  This results in a lower response and longer time delays.

\begin{figure*}
\centering
\includegraphics[width=\textwidth]{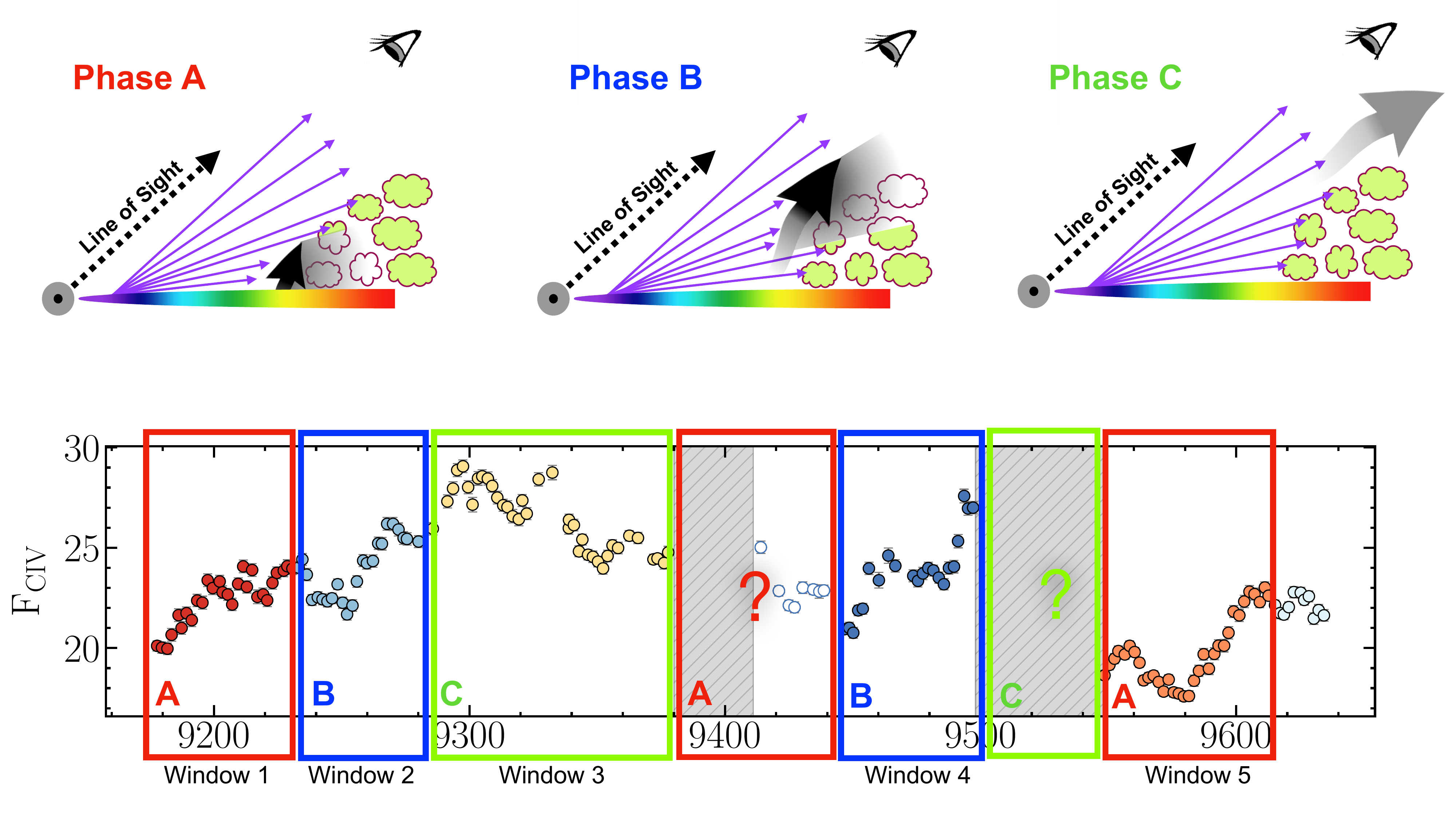}
 \caption{Schematic view of the BLR. The STORM2 campaign observed three phases of obscuration and outflow. The top panel shows these three main phases. In each phase, the BLR gas is illustrated by clouds. Those BLR clouds unobscured by the partially opaque, outflowing gas are shown with green clouds. By contrast, those clouds that lie within the shadow of the obscuration are illustrated with uncolored clouds.
 The thick arrow shows the obscuration, the ionizing continuum emission is identified by the collection of purple arrows, and the ever-present outflow is shown as a stream that is launched from the accretion disk. In Phase A, obscuration blocks the inner BLR from the continuum emission. Shorter lags disappear because it takes longer for the BLR gas at larger radii to realize that the continuum emission is blocked (windows~1 and 5). During Phase B, the obscuration flows upwards and outwards into our line of sight. The base of the wind becomes transparent so that the continuum emission once again reaches the inner edge of the BLR. During this time we ``see" short lags since only the inner BLR region reprocesses the continuum emission (Windows 2 and 4). Then, in Phase C, the obscuration at low elevations becomes transparent enough for the continuum emission to uniformly reach all of the BLR clouds, presumably the true extent of the BLR. The bottom panel shows the \ion{C}{4} light curve, and the five identified windows. The colored frames on each window of the light curve correspond to the matched-color phase. Phase A corresponds to windows 1 \& 5, Phase B corresponds to windows 2 \& 4, and window 3 corresponds to Phase C. The grey shadings, which illustrate the two HST safing incidents, could plausibly undergo a similar scenario. Although our schematic is two dimensional, we argue that the variations in the emission line require the obscuration to be azimuthally symmetric, as the observed changes in response encompass the entire BLR, not just our line of sight.}
\label{fig:blr_cartoon}
\end{figure*}

A thought experiment illustrates a plausible scenario that accounts for longer lags being associated with intervals of higher obscuration. The CCF for \ion{C}{4} shows significant response from the BLR over a range of 1 to 20 light days. If we suddenly interpose an opaque screen between the ionizing continuum and the BLR, it is the interior regions of the BLR with the shorter lags that will first notice the lack of continuum radiation and stop responding. More distant regions still ``see'' the radiation emitted before the screen was in place. The resulting lags we measure, therefore, will be weighted more heavily toward gas at greater distances with longer lags. As this change in obscuration propagates outward and the BLR adjusts to the new level of obscuration, lags will revert to more characteristic, presumably shorter timescales. In addition, the obscuring screen is likely decreasing in opacity, as evidenced by the different tracks in response shown in Figure~\ref{fig:lc_overlap} and by the changing levels of line-of-sight obscuration as measured by the broad UV absorption troughs (bottom panel in Figure~\ref{fig:lc_overlap}). This evolution in the line-of-sight obscuration we actually measure from lower values at the beginning of the campaign to a peak $\sim$30 days later may reflect the timescale for gas rising from the accretion disk in the equatorial zone where it obscures the BLR and outflowing to a height where it intercepts the line of sight. For an obscurer located at 1 light day near the inner edge of the BLR, the dynamical timescale for a central black hole mass of 3.85 $\times 10^7 M_{\odot}$ \citep{Bentz2015} is 30 days, consistent with the interpretation that the evolution of the line-of-sight obscuration is due to accretion disk material being lifted off the inner accretion disk and transported on a dynamical time scale to cross our line of sight. Figure~\ref{fig:blr_cartoon} illustrates the evolution of the obscuration as it blocks the inner broad line region from the continuum. We presume that prior to the beginning of the campaign there was no obscuration and the BLR was fully ionized. During Phase~A, the obscuration appears and the innermost BLR is shielded first. This causes the shortest lags to disappear. During Phase~B, the obscuration flows upwards and outwards into our line of sight over a timescale of $\sim$30~days, which is the dynamical timescale for material in the inner BLR. At this time, the base of the outflowing wind becomes transparent, and the inner BLR gas can ``see" the continuum. In Phase~C, the obscuration has lifted and is transparent everywhere, such that the entire BLR is illuminated by the continuum. We also note that as shown by \citet{Dehghanian2020}, the energy absorbed by the obscurer is re-radiated, but this re-emitted radiation is difficult to detect. It is mostly in the form of very broad line emission and diffuse continuum emission, and its intensity is much reduced since it is isotropically re-emitted.

To express these ideas more quantitatively, we follow the arguments by \cite{Goad2014}. AGN variability is well described by a damped random walk \citep{Czerny2003, Uttley2005, Kelly2009, Kozlowski2010, MacLeod2010, Zu2011}. On short timescales this variability has a power density spectrum that is a power law with spectral index $-2$.  For understanding the BLR response in \mrk, the essential feature is that short time-scale continuum fluctuations have less power and lower amplitude than those at longer timescales. When the ionizing continuum is heavily obscured, the amplitude of continuum fluctuations as seen by the BLR is significantly reduced.  Although amplitudes are suppressed on all timescales by the obscuration, the SNR we achieve in our observations limits our ability to detect some amplitudes and timescales.

The low-amplitude, short-timescale, rapid fluctuations that we see in the observable UV are suppressed in the ionizing continuum viewed by the BLR, up to as much as a factor of 10 for the 90\% covering fractions seen in the X-ray \citep{Partington2023}. This prediction of a reduced amplitude of fluctuations in the BLR is consistent with the  significantly lower amplitudes in the RMS spectrum for Window~1 as seen in the top right-hand panel of Figure~\ref{fig:mean_rms}. So, even though our flux measurement errors of better than 1.5\% allow us to easily see a response to 5—10\% fluctuations when the source is unobscured, these become undetectable during periods of heavy obscuration. During these periods we are only able to measure a response in the BLR for the stronger variations (50\% to 200\%) on longer timescales.  Following the simulations by \cite{Goad2004}, this favors recovering a CCF with longer lags.  However, note that a short timescale response is still present in the CCFs for temporal windows~1 to 5 (see the second panel in Figure \ref{fig:bimodal_lag}) even though it is not dominant. In particular, notice the secondary peak for Window~1 at the short lags typical of the lightly obscured Window~3.  During the more transparent phase, Window~3, the BLR is illuminated by the full ionizing continuum, with all the rapid fluctuations we see in the observable UV also present in the variations of the ionizing flux.  These more rapid variations are diminished at large radii by geometrical dilution, so the CCF is biased toward shorter timescales, and the peak shifts to short lags.

A similar process with the appearance of the injection of a new screen of opaque material into the outflow may have occurred at the end of the campaign (Window~6), which may be indicative of a non-responsive ``holiday" interval in the \ion{C}{4} flux during the last nine epochs of the HST campaign. This might be testable with the behavior of the H$\beta$ emission line as tracked by the on-going ground-based campaign.
\section{Summary and Future Work}\label{sec:summary}

The AGN STORM 2 campaign on \mrk\ reveals complexities in broad-line region structure and its response to continuum fluctuations that go beyond the simplest concepts originally envisioned for reverberation-mapping experiments. As noted in Paper~2, the emission-line light curves for \mrk\ are not merely smoothed, shifted, and scaled versions of the continuum light curve. In this paper we have shown that different temporal windows in the STORM 2 campaign have different responses to the continuum fluctuations, with the \ion{C}{4} emission line flux in each time interval showing a different response to continuum fluctuations, and a different time lag. These time lags range from 2—13 days. The different temporal windows correspond to significant variations in the properties of the obscuring gas. Temporal windows with the longest lags correspond to periods of increasing obscuration, with the obscurer shielding and diminishing the response of the innermost regions of the BLR. Temporal windows with the shortest lags occur in intervals with diminishing obscuration. The changing spectral energy distribution of the ionizing flux reaching the BLR may be responsible for the changes in line responses in the different temporal windows.

In future work, once we have modeled the effects of absorption on the emission-line profiles and deblended the adjacent emission lines, we will extend the analysis presented here to Ly$\alpha$, \ion{N}{5}, \ion{Si}{4}, and \ion{He}{2}.

\software{\texttt{PyCCF} \citep{Sun2018b}, \texttt{Scikit-learn} \citep{Pedregosa2011}.}


This paper is the fifth in a planned series of papers by the AGN STORM~2 collaboration. Our project began with the successful Cycle 28 HST proposal 16196 \citep{Peterson2020}. Support for Hubble Space
Telescope program GO-16196 was provided by NASA through
a grant from the Space Telescope Science Institute, which is operated by the Association of Universities for Research in Astronomy, Inc., under NASA contract NAS5-26555. We are grateful to the dedication of the Institute staff who worked hard to review and implement this program. We particularly thank the Program Coordinator, W. Januszewski, who made sure
the intensive monitoring schedule and coordination with other facilities continued successfully. 
\begin{acknowledgments}
Y.H. acknowledges support from NASA grant HST-GO-16196, and was also supported as an Eberly Research Fellow by the Eberly College of Science at the Pennsylvania State University. Research at UC Irvine has been supported by NSF grant AST-1907290. Research at Wayne State University was supported by NSF grant AST 1909199, and NASA grants 80NSSC21K1935 and 80NSSC22K0089. E.K. acknowledges support from NASA grants 80NSSC22K0570 and GO1-22116X. N.A. acknowledge support from NSF grant AST 2106249, as well as
NASA STScI HST grants AR-15786, AR-16600, AR-16601, and AR-
17556. H.L. acknowledges a Daphne Jackson Fellowship sponsored by the Science and Technology Facilities Council (STFC), UK. M.C.B. gratefully acknowledges support from the NSF through grant AST-2009230. T.T. and P.R.W. acknowledge support by NASA through grant HST-GO-16196, by NSF through grant NSF-AST 1907208, and by the Packard Foundation through a Packard Research Fellowship to T.T. G.J.F. and M.D. acknowledge support by NSF (1816537, 1910687), NASA (ATP 17-ATP17-0141, 19-ATP19-0188), and STScI (HST-AR- 15018 and HST-GO-16196.003-A). P.B.H. is supported by NSERC grant 2017-05983. M.V. gratefully acknowledges financial support from the Independent Research Fund Denmark via grant number DFF 8021-00130. D.H.G.B. acknowledges CONACYT support \#319800 and of the researchers program for Mexico. D.C. acknowledges support by the ISF (2398/19) and the D.F.G. (CH71-34-3). A.V.F. was supported by the U.C. Berkeley Miller Institute of Basic Research in Science (where he was a Miller Senior Fellow), the Christopher R. Redlich Fund, and numerous individual donors. The UCSC team is supported in part by the Gordon and Betty Moore Foundation, the Heising-Simons Foundation, and by a fellowship from the David and Lucile Packard Foundation to R.J.F. 
D.I., A.B.K and L.Č.P. acknowledge funding provided by the University of Belgrade--Faculty of Mathematics (contract No. 451-03-47/2023-01/200104) and Astronomical Observatory Belgrade (contract No. 451-03-47/2023-01/200002) through the grants by the Ministry of Science, Technological Development and Innovation of the Republic of Serbia. D.I. acknowledges the support of the Alexander von Humboldt Foundation. A.B.K. and L.Č.P. thank the support by Chinese Academy of Sciences President’s International Fellowship Initiative (PIFI) for visiting scientist.
P.D. acknowledges financial support from NSFC grants NSFC-12022301, 11873048, and 11991051.
\end{acknowledgments}

\bibliography{main.bib}

\end{CJK*}

\appendix
\section{Considerations for Temporal Window Selections}
\label{sec:appendix}

To test for possible time interval edge effects, given anticipated \ion{C}{4} time delays, we extend the length of each time interval. We modify the start date of $F_{\rm 1180}$ continuum points to start $10$~days earlier and also extend the end dates on the \ion{C}{4} light curves to be 10~days later than the dates reported in Table \ref{tab:table1}. The reported lag measurements in Table~\ref{tab:table2} are the result of this extended continuum search range. We find that these extended boundaries result in time delays that are consistent within $1\sigma$ with the exact time interval ranges reported in Table \ref{tab:table1}.

We also note that the data immediately after the first HST safing event are more sparsely sampled for $\sim$30 days and have a longer mean cadence of 4 days (compared to the expected 2-day cadence). This raises two issues. First, since the continuum light curve is driving the emission-line response, the absence of continuum data during the safing event makes it difficult to measure a response for observations immediately following the safing interval.
Second, the larger intervals between the observations immediately after the safing event degrade the resolution of the time-delay measurement. Therefore, we exclude THJD = 9413-9445 from this window, ensuring sufficient sampling and giving some continuum coverage post-safing. Thus we only consider THJD = 9445$-$9498 for Window~4 (blue). We do not face the same issue following the second safing event, since the light curve sampling is uniform right after the safing incident and consistent with the expected 2-day cadence. 

Near the end of the campaign, after the last continuum peak at THJD$\approx$9600 (see Figure~\ref{fig:lc_windows}), the BLR response to the  diminishing continuum is significantly reduced, where the continuum falls by a factor of 2 in 20 days, but the line fluxes fall by only a few percent.
Although the continuum flux is rising after THJD = 9615, only \Lya\ shows a clear response.
All other emission lines, including \ion{C}{4}, remain less responsive, suggesting that the BLR may have entered another holiday period at the end of the campaign. This decorrelation continues for the remaining 20 days of the campaign. We therefore end Window~5 at THJD = 9615 (orange), and we do not carry out a detailed time-delay analysis of Window~6 (light blue).

In summary, the periods that we remove from time-delay measurements are:
\begin{enumerate}
    \item THJD = 9413$-$9445, due to sparse sampling (mean of 4 days) immediately after an extended HST safing gap.
    \item THJD = 9615$-$9634, due to the likelihood of entering a BLR holiday state \citep{Goad2016}.
\end{enumerate}
  
\end{document}